\documentclass[aps,prb,twocolumn,superscriptaddress,showpacs,showkeys,amsfonts,amssymb,amsmath]{revtex4}
\usepackage{graphicx}
\usepackage{ulem}
\usepackage[colorlinks=true,linkcolor=blue,citecolor=blue,urlcolor=blue]{hyperref}
\usepackage{float}
\usepackage{xspace}
\usepackage[T1]{fontenc}

\newcommand{\lasr}{La$_{1-x}$Sr$_x$MnO$_3$}
\newcommand{\la}{La$_{7/8}$Sr$_{1/8}$MnO$_3$}
\newcommand{\lacuo}{La$_{2-x}$Sr$_x$CuO$_4$}

\begin{document}

\author{Martine Hennion*}
\affiliation{Laboratoire L\'eon Brillouin, Universit\'e Paris-Saclay, CNRS, CEA, F-91191 Gif-sur-Yvette, France}
\author{Alexandre Ivanov*}
\affiliation{Institut Laue-Langevin, 71 avenue des Martyrs, CS 20156 F-38042 Grenoble, France}
\author{Claudine Lacroix}
\affiliation{Institut N\'eel Grenoble-Alpes CNRS, 25 avenue des Martyrs, BP 166  F-38042 Grenoble, France}
\author{Bernard Hennion}
\affiliation{Laboratoire L\'eon Brillouin, Universit\'e Paris-Saclay, CNRS, CEA, F-91191 Gif-sur-Yvette, France}
\title{ From chessboard of bipolarons of size 4a in cubic \la\ 
 \\ to stripes of the same bipolarons in layered high $T_c$ cuprates}

\date{\today}
\begin{center}

\end{center}

\begin{abstract}

*corresponding authors: martine.hennion02@gmail.com and aivanov@ill.fr\\
\newline

The compound \lasr\ exhibits a charge order (CO) state at $x\approx 1/8$ and $T<T_{co}$, which recalls the CO state with a decrease in the temperature of the superconducting transition, $T_c$, observed in all cuprates at this doping value. Local excitations of lattice and magnetic origins measured in the two-dimensional metallic state of \la\ reveal the existence of bipolarons of size $4a$ resulting from structural and antiferromagnetic pairings of hole-rich orbital polarons of size $2a$. They are intertwined with hole-poor domains in a disordered state at $T>T_{co}$ which become ordered on a chessboard organized in a 3D-order state of ferromagnetically paired polarons at $T<T_{co}$. Applied to the CuO$_2$ planes of the cuprates of the "214" family, this model produces stripes of bipolarons intertwined with stripes of antiferromagnetically arranged spins, hole-poor, both of size $4a$, leading to a spin density wave with a wave vector $\delta=1/8$, a charge density wave with $q=1/4$, the Yamada laws $\delta(x)=x$ and $T_c\propto \delta$ and a decrease of $T_c$ at x=1/8. This work invokes relevance of a bipolaronic origin of high $T_c$ superconductivity, in which bipolarons of size $4a$ can play a major role.

\end{abstract}

%
\maketitle

{\bf I Introduction } \\ 

Among manganites, the pseudocubic \lasr\ (edge of a cubic lattice unit cell $a$=3.9\r{A}) appears particularly interesting because its phase diagram shows characteristics typical of cuprates with high superconducting (SC) temperature $T_c$ (see the phase diagrams in the Supplementary Material (SM) of this paper, Ref.\onlinecite{Supp-Mat}, Fig.SM-1).
In the doping range $0.1 \le  x \le 0.15$ around $x=1/8$ and for $T<T_{co}$\cite{Paraskelvopoulos00}, this material presents a three-dimensional (3D) charge order state without a consensus reached to date on its origin\cite{Argyriou96,Yamada96,Endoh99,Yamada00,Tsuda01,Cox01,Geck04,Wei10}. A previous study of spin dynamics has shown that the charge order (CO) state arises from a ferromagnetic (F) and metallic state with a two-dimensional (2D) character\cite{Hennion06}. This 2D character is the consequence of the magnetic structure of LaMnO$_3$, which consists of ferromagnetic planes weakly coupled by an antiferromagnetic coupling (AF) in their perpendicular direction\cite{Pinsard97,Hennion06}. This 2D metallic state stabilizes at a higher temperature $T_{FM}$,
where the charge-induced magnetic coupling is equivalent along the two directions of the MnO bond\cite{Hennion06}. Specifically for this compound with $x$=1/8, the values $T_{FM}$=181K and $T_{co}$=159K have been determined\cite{Hennion05}. 

In the same doping range of the phase diagram of all high-$T_c$ cuprates, a depression of superconducting temperature $T_c$(x) is observed\cite{Kofu09,Keimer15}. At the same time, a long range charge density wave ($CDW$)\cite{Comin16,Wen19} and a spin density wave ($SDW$) with periods $\approx 4a$ and $\approx 8a$, respectively, occur in
La$_{2-x}$Ba$_x$CuO$_4$\cite{Tranquada95,Tranquada08} and La$_{2-x}$Sr$_x$CuO$_4$\cite{Yamada98} with optimal order at $x=1/8$. Several models have been proposed\cite{Zaanen89,Tranquada95,White98,Tranquada04,Berg07,Christensen07,Tranquada13,Agterberg20} while the microscopic origin of high $T_c$ superconductivity remains uncovered and the role of $CDW$ which competes with superconductivity is still debated\cite{Keimer15,Uchida21}. 

In the present paper, we report new additional phonon acoustic branches in \la, at T<$T_{co}$ and T>$T_{co}$ along the MnO bond directions that lie in restricted ranges of wave vector $q$ in the reciprocal space. Based on a previous study at $x=0.2$ (Refs.\onlinecite{Hennion19,Supp-Mat}) and considering previous data on magnetic excitations along several symmetry directions (Refs.\onlinecite{Hennion06,Hennion19}), they can be attributed to hole-rich domains of size $4a$ intertwined with hole-poor domains of the same size organized in a chessboard structure at $T<T_{co}$. They coexist with static charge density waves  (CDW) of period $4a$ ($4a\sqrt 2$) along [100] ([110]) identified from gaps in the magnetic dispersion law $Dq^2$. 
 The bipolaronic origin of the hole-rich domains appears in the 2D metallic state, where the chessboard is expected to be distorted.
 We recall that a bipolaron results from the structural and AF pairings of two polarons which are defined in real space by the coupling of one charge with phonons. In the present work, these pairing properties are considered for bipolarons with a hole-rich character that are separated by hole-poor domains.  Applying this picture to the layered structure of high $T_c$ cuprates which shares the same orbital structure, one gets a picture of stripes of bipolarons of size $4a$ intertwined with stripes of spins arranged antiferromagnetically, hole-poor. A superconducting state is expected at $T_c$ from the band created by their fluctuations. At $x=1/8$ as the size of the hole-poor domains between the stripes of bipolarons decreasing with $x$ becomes equal to the size $4a$ of the bipolarons, the CDW that coexists with the bipolarons acquires a long-range period $4a$ that produces partial immobilization and ordering of the stripes of bipolarons and, accordingly, a decrease in the value of $T_c$. 

The origin of the additional branches of the acoustic phonon was first discussed for $x=0.2$ in \lasr\cite{Hennion19}. There, three acoustic phonon branches with transverse and longitudinal character labeled TA*(q), TA$^{perp}$(q) and LA*(q) have been observed in restricted ranges of $q$ values along [100] + [010] + [001] (see TA* and TA$^{perp}$ in Fig.SM-2a of Ref.\onlinecite{Supp-Mat} adapted from Ref.\onlinecite{Hennion19}). They coexist with the longitudinal LA(q) and transverse TA(q) phonon branches of the pseudocubic structure common to the twinned domains. The wave vector {\bf q} along [100] + [010] + [001], superposed by twining, corresponds to waves propagating along the MnO bond directions and $q$ is expressed in reciprocal lattice units (rlu) of the cubic structure so that {\bf q} = ($2\pi/a$)[q00] {\it etc} ($a$=3.9\r{A}).

In the problem of charges subjected to multiple competing interactions, the spectrum of acoustic phonon branches may indicate that there exists an effective medium characterized by the LA and TA branches of the pseudocubic structure with the highest intensity, in which hole-rich domains characterized by local excitations appear as additional branches in the restricted range [$q_{min}-0.5]$ along the MnO bond directions. The lower limit $q_{min}$ of the range of $q$ can be used to determine the size of the domains if the spatial distribution of these domains preserves the phase of the excitations. This property is obtained if the domains are in contact along some direction or if they are intertwined with a hole-poor domain of the same size that transmits the phase along some other direction. Both situations are observed in our data at $x=0.2$ and $x=1/8$.
These branches are observable if the lifetime of the domain is long compared to the reciprocal frequency of the local phonon excitations, so they are no longer observed in the true metallic compound with $x=0.3$\cite{Reischardt99}. We get a weakly inhomogeneous picture with localized (hole rich) and non-localized (hole poor) components for the charge, less accentuated than in the images of charge segregation or phase separation previously used\cite{Hennion19}.

At $x=0.2$, the value $q_{min}=0.25$ of the additional LA* and TA* branches has determined the linear size $2a$ of the "hole rich" domains using the relation $\xi={\pi}/|{\bf q}_{min}$| so that the boundary of the Brillouin zone $q_{min}$ = 0.5 rlu corresponds to the smallest distance of coupling $\xi=a$. The maximum intensity observed at $q_{min} = 0.25$ rlu in the TA* branch is a consequence of the stationary character of the local vibrations of size $2a$ in contact, defining chains along the MnO bond direction\cite{Aubry}. The size $2a$ corresponds to the ferromagnetic "orbital polaron" introduced to explain the charge-ordered states of manganites ($x=0.25, x=0.5$) in which the charge-phonon coupling is neglected so that polarons or domains of hole-density do not appear\cite{Kilian99,Mizokawa00}. In such an ionic picture, the $Mn^{3+}$ ion consists of an unoccupied external level showing a doubly degenerate orbital state $e_g$ (${x^2-y^2}$, $3z^2-r^2$) well separated in energy from the inert spin core $S=3/2$\cite{Edwards02}. Whatever its 1D, 2D or 3D character, the orbital polaron has been defined by the orbital states $T^x$, $T^y$, $T^z$ ($T^{\alpha}=3\alpha^2-r^2$, $\alpha=x,y,z$) of the $Mn^{3+}$ sites with lobes pointing toward the common neighboring site ($Mn^{4+}$) that carries the hole\cite{Kilian99,Mizokawa00}. At $x=0.2$, the complete superstructure of chains of 1D polarons was determined considering the other transverse acoustic branch $TA^{perp}(q)$ in the perpendicular direction superposed by twining. In contrast to TA* and LA*, this branch is observed in the small range of q values up to the cutoff point $q_c\approx 0.35$ observed with an energy far from TA* by $\Delta=3 meV$. This cutoff not only determines the periodic distribution $2\pi/q_c=3a$ of the chains but can also be interpreted as the zone boundary of a fictitious Brillouin zone defined during the short lifetime of the chain direction. In this way, the excitation at $q_c\approx 0.35$ corresponds to the transverse excitations in phase opposition of the hole-rich and hole-poor domains, intertwined, each with a rigid thickness $1.5a$. This thickness value suggests that the hole-rich density extends along the direction perpendicular to the chains or that the $p$ and $d$ orbitals are hybridized. 
One gets a picture of chains of polarons of thickness $1.5a$ with periodic distance $3a$ ($x\approx 0.17$) similar to an electronic crystal liquid\cite{Kivelson98} (see the sketch in Fig.SM-2b of Ref.\onlinecite{Supp-Mat}). The same periodic structure of charges appears in the magnetic excitations, as shown below.

We recall that in metallic manganites, along the directions of charge hopping (or tunneling), the magnetic spectrum consists of two distinct parts, depending on the range of q values. A law $Dq^2$ appears at small values of $q$ related to spin waves of the electronic band origin\cite{Edwards02}, and a discrete energy spectrum spreads at larger values of $q$. This discrete energy spectrum can be explained by considering that the hopping (or tunneling) of the charges has two effects that can be described in an ionic picture as follows: i) it induces a local ferromagnetic coupling between a $Mn^{3+}$ spin and its $Mn^{4+}$ neighbor\cite{Zener51}, ii) it induces a fluctuation of the anisotropic $e_g$ orbital state of the $Mn^{3+}$ sites, first neighbor of the $Mn^{4+}$ sites as the position of the charge varies. The atomic structure being determined by the orbital states, their fluctuations should be strongly coupled to phonons. At low doping values such as $x=1/8$ and x=$0.2$ where the charges are strongly correlated, their collective motion induces orbital fluctuations that are also strongly correlated so that they appear in the low energy range as q-dependent phonon branches. Their magnetic character\cite{Moussa03,Hennion19} is the consequence of the coupling between the orbital and spin degrees of freedom at the $Mn^{3+}$ sites\cite{Van-den-Brink98,Kugel73,Tokura00}.

At $x=0.2$, the absence of a law $Dq^2$ in the magnetic spectrum along [111] indicates that the hopping (or tunneling) of the charges is restricted to the MnO bond directions (see Fig.SM-4 of Ref.\onlinecite{Supp-Mat}). 
Along these directions, the discrete energy spectrum observed beyond the law $Dq^2$ is divided into two sets of E(q) branches, lying at $E<27$ meV with $q\ge q_{min}\approx 0.35$ rlu and at $E>27$ meV with $q\ge q_{min}=0.25$ rlu (see 
Fig.SM-3a of Ref.\onlinecite{Supp-Mat}). Since these characteristic values $q$ are the same as those of the two additional transverse phonon branches $TA^*$ and $TA^{perp}(q)$, separated in energy by the same value $\Delta=3 meV$ at $q\approx 0.35$, we conclude that the discrete magnetic spectrum arises from the motion or fluctuation of the same superstructure of the hole-rich and hole-poor domains. The two sets of branches E (q) are therefore attributed, for $E>27 meV$ and $q_{min}=0.25$ rlu, to the 1D orbital polarons (hole-rich domains) of size $2a$ in contact along chains, and, for $E<27 meV$ and $q_{min}\approx 0.35$, to the "hole-poor" domains intertwined with the chains along the perpendicular direction, defining the same thickness $1.5 a$ of the hole-poor domains as the branch $TA^{perp}(q)$.
A fully coherent picture appears by considering that, in the upper energy range ($E>27 meV$), the set of three branches agrees with the three expected MnO bond directions of the chains. At $T<T_{FM}$ their magnetic energy that coexists with the phonon energies starts to increase and becomes connected to the law $Dq^2$, highlighting the 1D character of the metallic and ferromagnetic state during the lifetime of the chain direction (see Fig.SM-3 in Ref.\onlinecite{Supp-Mat} adapted from Ref.\onlinecite{Hennion19}). 
\\

\begin{figure}[t]
\includegraphics[width=8cm]{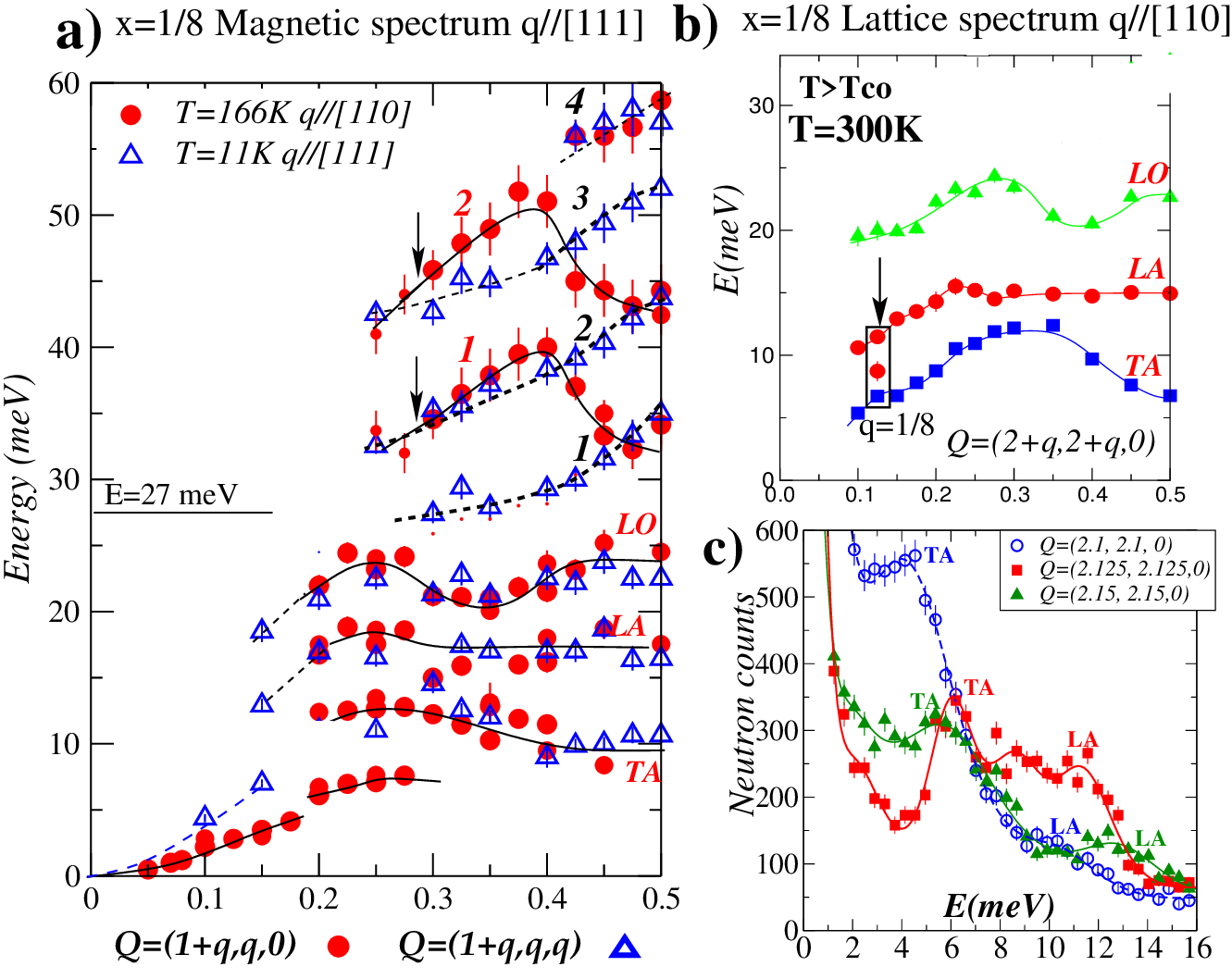}
\caption{\label{figure1} 
{\bf a)} Magnetic spectrum along [110] at T$>T_{co}$ (full red circles) and along [111] at T$<T_{co}$ (empty blue triangles). At $T>T_{co}$ we have checked that the [110] direction is equivalent to the [111] direction which was used in the highest energy range to open the available energy window. The two (four) branches observed along [110] ([111]) arise from AF paired (F paired) hole-rich domains fluctuating between the diagonals of squares (cubes). {\bf b)} Phonon spectrum determined at $T=300K$ along [110]. The rectangle outlines the additional excitation observed between LA and TA at q=0.125 rlu. The anomalous increase of their intensity is shown in the raw data of panel {\bf c)}.  In all excitation spectra, the vertical lines are error bars and the continuous and dashed lines are guides to the eye.}

\end{figure}

\begin{figure}[t]
\includegraphics[width=8cm]{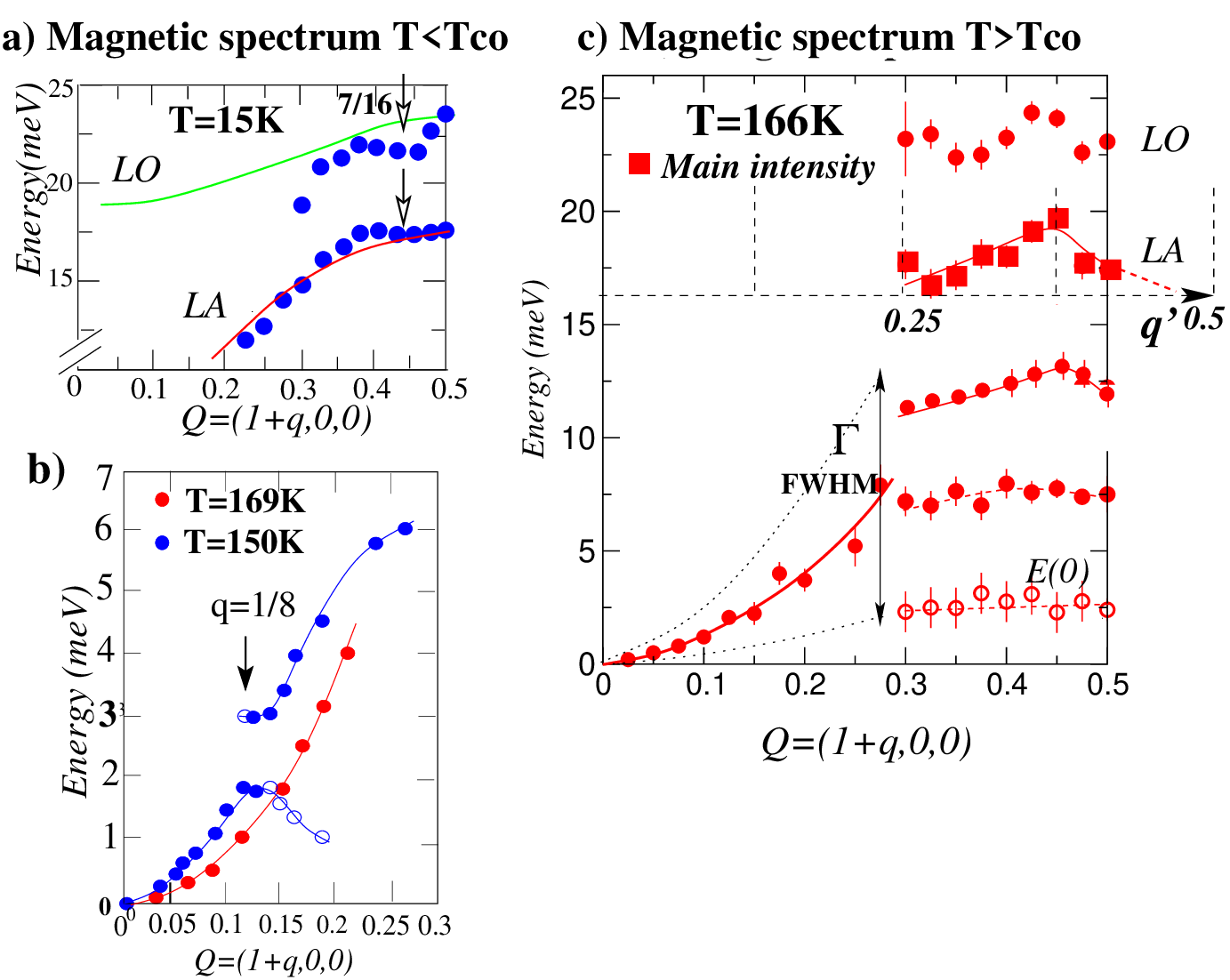}
\caption{\label{figure2} Magnetic spectrum along the MnO bond directions. {\bf a)} at $T<T_{co}$ E(q) branches of the magnetic spectrum adapted from Ref.\onlinecite{Hennion05} showing a dip at $q=7/16$ rlu in the $E=LA$ and $E=LO$ branches corresponding to the "in-plane" magnetic fluctuations. The upper energy branch occurring at $E=LO'$ at $T<T_{co}$ has been omitted for clarity. {\bf b)} Quadratic law $Dq^2$ measured at T=150K ( blue circles) adapted from Ref.\onlinecite{Hennion06} showing an energy gap at $q=1/8$ rlu outlined by an arrow and compared to the $Dq^2$ measured at T=169K (red circles). {\bf c)} Maximum of energy observed at $q=0.45$ rlu in the upper energy branches of the magnetic fluctuations. The $q'$ scale is defined by $q'=(0.83)^{-1}q$ (see the text). The E(0) branch, also observed in lattice excitations, but not displayed, can be attributed to a binding energy of the bipolaron to the lattice\cite{Hennion19}. $\Gamma$ is the energy line-width. }
\end{figure}

\begin{figure}[h]
\includegraphics[width=8cm]{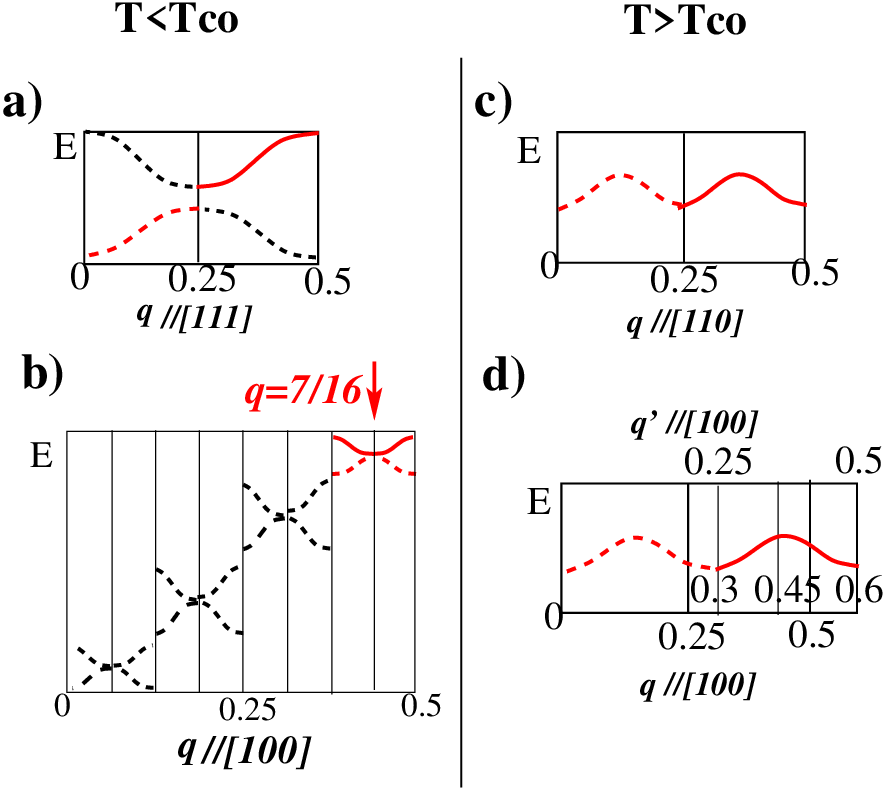}
\caption{\label{figure3} The dispersion branches sketched by continuous lines reveal four distinct experimental situations. {\bf a)} an F coupling at the interface between two neighboring F domains, each of size $a\sqrt{3}$ along [111], {\bf b)} a F coupling between adjacent F domains of size $2a$ showing a period $8a$. {\bf c)} a AF coupling at the interface between two neighboring F domains, each of size $a\sqrt{2}$ along [110] {\bf d)} an AF coupling at the interface between two neighbor F domains of size $2a'$ with $a'=0.25/0.3 a$ and $q'$=$q$ x $0.3/0.25$ along the MnO directions of the planes (see the text).}
\end{figure} 

\ \  \

{\bf II Results: Magnetic and lattice excitations at x=1/8}.

In Section II-1 the previously reported magnetic spectra are interpreted in the hole-rich, hole-poor model\cite{Hennion06}, in Section II-2, we present new lattice excitations obtained at $T<T_{co}$ and in Section II-3, new lattice excitations obtained at $T>T_{co}$. In the discussion (Section III), we show how the existence of bipolarons of size $4a$ can be a key to interpreting some properties of the layered high $T_c$ cuprates, especially those with the simplest structure of the 214 family.\\   

{\bf 1- Magnetic excitations in the 3D ferromagnetic state ($T<T_{co}$) and in the 2D metallic state ($T>T_{co}$) in the hole-rich, hole-poor model }\\
In the magnetic spectrum, the new situation at $x=1/8$ with respect to $x=0.2$, is the existence of a law $Dq^2$ along several directions of symmetry. Moreover, the observation of additional acoustic phonon branches in the $[q_{min}-0.5]$ range of the $q$ value at $x=1/8$ as at $x=0.2$ (they are described in the next Section) allows us to interpret the discrete energy magnetic spectrum in terms of domains in direct space.
 
Fig.\ref{figure1}-a presents the experimental data along the "diagonal" directions, [111] ($T<T_{co}$ or 3D ferromagnetic state) and [110] ($T>T_{co}$ or 2D metallic state), complemented and improved over the previous report\cite{Hennion06} whereas Fig.\ref{figure2}-c presents those along the MnO bond directions\cite{Moussa03,Hennion06}. Along the diagonal directions, the magnetic spectrum is divided into two sets of branches E(q) $E<27meV$ and $E>27meV$. From our study at $x=0.2$, they are readily attributed to the magnetic excitations of the hole-poor ($E<27$ meV) and hole-rich ($E>27$ meV) domains. This separation in two energy ranges indicates that the excitations of the two types of domain do not interfere or that the hole rich and the hole poor domains are in contact along distinct "diagonal" directions. As expected, the low-energy branches attributed to hole-poor domains lie approximately in coincidence with the acoustic branches reported in Fig.\ref{figure1}-b.

In contrast, along the MnO bond directions reported in Fig.\ref{figure2}-c, a single set of branches is observed. There, the two types of excitation interfere or, equivalently, the two types of domain are intertwined.
 
The complete characterization of the hole-rich domains is provided by the value $q_{min}$ and the dispersion with $q$ of the branches. They are described in the 3D charge order ($T<T_{co}$) and the 2D metallic ($T>T_{co}$) states successively. 

a) In the 3D charge order state ($T<T_{co}$), 
i) along the diagonal [111], four branches, numbered from one to four in the figure, exhibit the value $q_{min}$=0.25 rlu (the higher energy branch being poorly defined).
Their dispersion with q reveals inflection points at $q=0.375$ rlu, middle of the range [$q_{min}$-0.5]. This dispersion
provides evidence of an F coupling at the interface between two adjacent F domains of size $2a\sqrt{3}$. They reveal domains of size $4a\sqrt{3}$, obtained by a pairing F of small domains $2a\sqrt{3}$ in contact along the four diagonals [111].  The dispersion with $q$ expected for a long-range order of such F-paired domains in contact along [111] is obtained by a folding of the Brillouin zone at $q=0.25$ as sketched in Fig.\ref{figure3}-a.
ii) Along the MnO bond directions where the excitations of the hole-rich and hole-poor domains interfere or are intertwined, the determination of the $q_{min}$ value is masked by the spectacular coincidence of the magnetic spectrum successively with the TA, LA, LO, and LO' phonon branches for $q\ge 0.25$ (see Fig.4 of Ref.\onlinecite{Moussa03}). However, the temperature evolution of the branches allows us to determine the value of $q_{min}$ = 0.25 rlu along [100] and [110], at the center of the large energy gap observed along these two directions where the intensity of the E(TA) branch jumps on the E(LA) one (see Fig.4-c of Ref.\onlinecite{Moussa03} and Fig.SM-5 of Ref. \onlinecite{Supp-Mat}). The most surprising feature is the departure of the ferromagnetic branches from the acoustic LA(q) and LO(q) branches observed at $q\approx 0.44=7/16$ along the MnO bond directions as seen in Fig.\ref{figure2}-a (adapted from 
Ref.\onlinecite{Hennion05}, improved from Fig.4 of Ref.\onlinecite{Moussa03}). Despite twinning, these two branches can be attributed to the directions [100] and [010] of the ($a, b$) planes (fluctuations of $T^x$ and $T^y$) since they still exist in the 2D metallic state, and, for $T<T_{co}$ (3D charge order state), they are separated in energy from the new branch with $E=LO'$ which occurs at $T<T_{co}$ readily assigned to the direction [001] (fluctuations of $T^z$). This upper energy branch has been omitted in Fig.\ref{figure2}-a for clarity. Using the same analysis as along [111], this dip value in the dispersion is attributed to a folding by eight of the Brillouin zone or a period $8a$ for the domains lying along the two MnO directions of the planes (see the sketch in Fig.\ref{figure3}-b).

\begin{figure}[t]
\includegraphics[width=8cm]{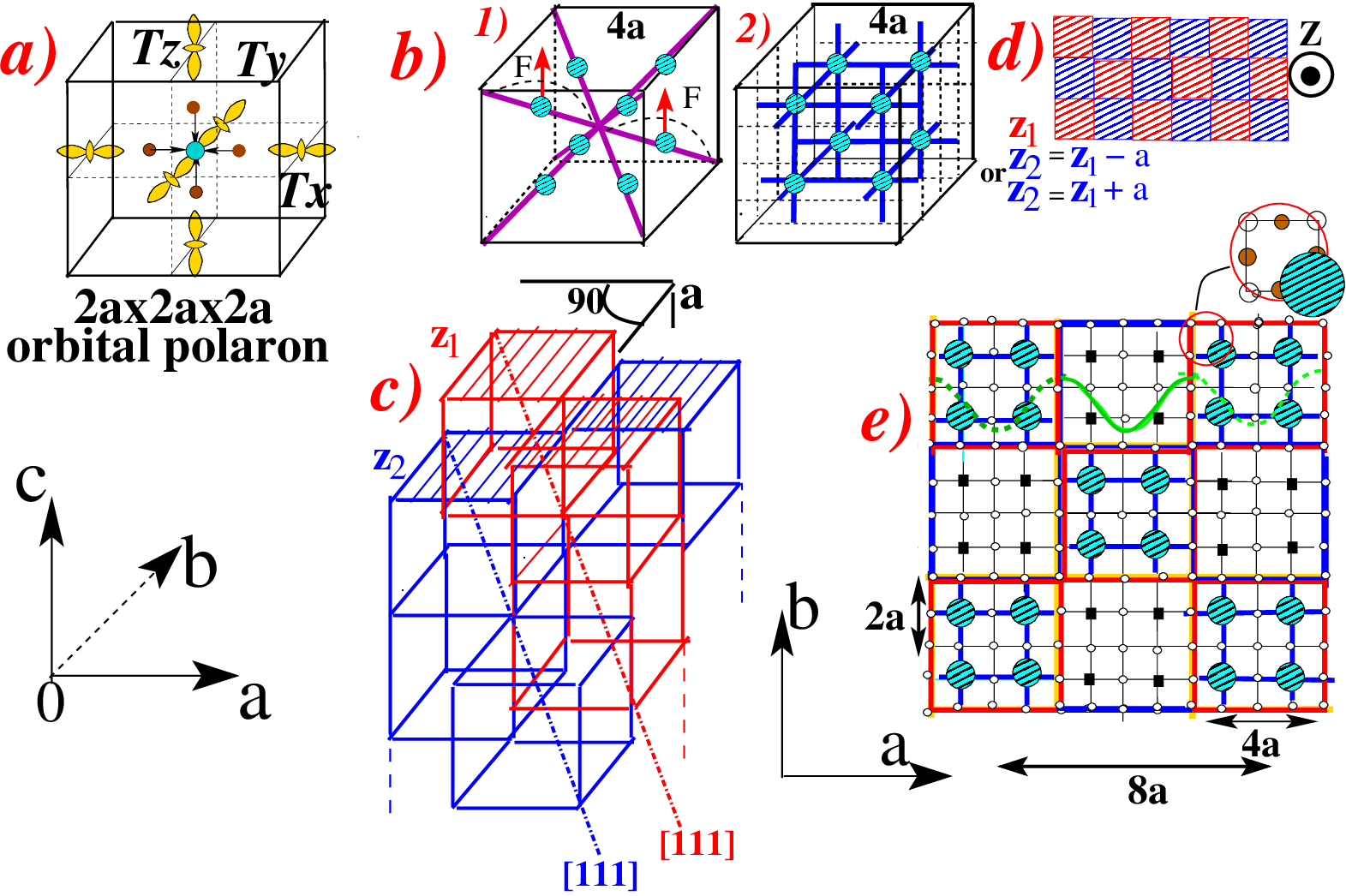}
\caption{\label{figure4} {\bf  3D superstructure of orbital bipolarons}. 
{\bf a)}: 3D ferromagnetic orbital polaron $2a$x$2a$x$2a$ defined by the orbitals $T^x$, $T^y$ and $T^z$ lying at the center of each face and pointing to the $Mn^{4+}$ site in an ionic picture (cyan circle), adapted from Ref.\onlinecite{Mizokawa00}. b) Cubes of side $4a$ containing eight charges (octopolarons) defined 1) from magnetic excitations that reveal F domains on the scale $4a\sqrt 3$ along the four [111] directions resulting from the F pairing of domains of size $2a\sqrt 3$ along these directions. They are observed thanks to orbital fluctuations that occur during their motion 2) from the additional lattice excitations that reveal bipolaronic domains of size $4a$ along the 3 MnO directions (12 links) observed during their short lifetime. c) 3D charge order defined by two families of columns of identical cubes of side $4a$ sketched with red and blue colors corresponding to a translation by $+a$ or $-a$ along $c$ axis.
Each cube is connected to cubes of same family by its four diagonals [111] (see the two dotted-dashed lines along one diagonal [111] of each family). 
d) Projection in (a, b) planes of the two families of cubes with bases respectively at the coordinates $z_1$ and $z_2=z_1-a$ or $z_2=z_1+a$. e) Chessboard of hole-rich and hole-poor domains of size $4a$ with alternation along the $\bf c$ axis determined by the 3D charge order. The white circles indicate  Mn sites. The large blue circles visualize the center of the hole-rich polarons in a given MnO plane. The green line indicates a CDW of period $4a$ along one MnO bond direction of the planes intertwined with the chessboard of bipolarons also inferred from our experiments. The CDW of period $4a\sqrt 2$ also observed along the [110] direction is not shown. The large red circle enhances the small red one. The charge density extends up to the oxygen sites sketched by filled brown circles.}
\end{figure}

From observations along these two symmetry directions, we conclude that the 3D space is filled by large cubes $4a$x$4a$x$4a$ obtained by ferromagnetic pairing of adjacent orbital polarons $2a$x$2a$x$2a$ along their four [111] diagonals (see Fig.\ref{figure4}-a,b). All these cubes are identical. There exists a unique arrangement of these large cubes that reconciles the fact that their diagonals are in contact along the four [111] directions, and, at the same time, their sides along the two MnO directions of the ($a, b$) planes produce a period $8a$. As sketched in Fig.\ref{figure4}-c, these two properties are realized by considering two families of columns of cubes $4a$x$4a$x$4a$. 
One family of cubes differs from the other one by a translation by $a$ or $-a$ along the {\bf c} axis. In the MnO planes, this 3D arrangement determines a chessboard of hole-rich and hole-poor domains of side $4a$ with an alternation of the two types of domains along the {\bf c} axis (see Fig.\ref{figure4}d,e). These hole-rich and hole-poor domains are represented by different squares of side $4a$ in the Fig.\ref{figure4}d,e and correspond to different cross-sections of these cubes by the planes MnO: in the middle of the cube $2a$x$2a$x$2a$ on the Fig.\ref{figure4}a and shifted up or down by the lattice spacing $a$ (a distance between adjacent planes MnO).

This superstructure of charges can be seen as a structure of "octopolarons" with one charge per 8 Mn sites ($x=1/8$). It provides the origin of the two static peaks previously reported\cite{Hennion05} that indicate the periods $2a$ (magnetic) and $4a$ (nuclear) along the {\bf c} axis\cite{Yamada96,Hennion05}.   

Moreover, gaps have been reported at $q=0.125$ rlu which open in the dispersion law $Dq^2$ for ferromagnetic magnons of metallic origin for ferromagnetic excitations, just below $T_{co}$ along the two symmetry directions [100] and [110] of the planes (see Fig.\ref{figure2}-b along [100] adapted from Ref.\onlinecite{Hennion06}). We attribute these gaps to new static spin density waves of period $4a$ (and $4a\sqrt 2$) along [100] (and [110]) which, in their turn, manifest appearance of charge density waves (CDW) with the same period polarized by the ferromagnetic spin ordered structure of the isolating charge-ordered low temperature phase. These charge density waves are intertwined with the chessboard of hole-rich/hole-poor domains (see the green line sketched for clarity along one MnO direction in Fig.\ref{figure4}-e). 
 
b) In the 2D metallic state ($T>T_{co}$), a strong change appears in dispersion with $q$ of the branches of the hole-rich domains. Moreover, the effect of disorder differs depending on whether the domains considered are along the diagonals [110] or along the MnO bond directions of the ($a, b$) planes. i) Along the diagonal directions [110], two branches numbered 1 and 2 in Fig.\ref{figure1} have been determined for the hole-rich domains (E>27 meV). A first decrease in intensity is observed at $q=0.3$ rlu (see the arrows) that prevents one from accurately determining $q_{min}$ ($\approx  0.25$ rlu). The modulation of the two branches exhibits a rounded shape with the maximum energy lying at $q\approx 0.375$ rlu in the middle of the range [$q_{min}$-0.5] (see raw data in Fig.SM-6 of Ref.\onlinecite{Supp-Mat}). This maximum corresponds to a double-scale AF coupling between the AF spin characteristics of an AF state. They provide evidence of an AF coupling at the interface between two adjacent F domains of size $\approx 2a\sqrt{2}$ and, by the way, domains of size $\approx 4a\sqrt{2}$ distributed with some disorder along the two [110] directions. The dispersion with q expected for a long-range order of AF-paired domains in contact along [110] is sketched in Fig.\ref{figure3}-c. 

ii) In contrast, along the MnO bond directions, the value $q_{min}$ = 0.3 rlu is clearly defined in Fig.\ref{figure2}-c and the modulation of the magnetic branch with the main intensity accurately determines the maximum energy at $q=0.45$ rlu at the middle of the range q=$[0.3-0.6]$. This value agrees with an AF pairing between to adjacent domains of incommensurate size $2a'$ observed in a fictitious Brillouin zone with $a'$ = 0.83 $a$ so that the zone boundary appears at $q=0.6$ instead of $q=0.5$ (see Fig.\ref{figure3}-d and note\cite{Note}). Such a large distortion of the lattice being unphysical, we recall that along this direction the excitations of the hole-rich domains interfere with those of the hole-poor domains. This discommensuration is therefore attributed to the varying thickness values of the intertwined hole-poor domains, lattice-locked, while the size of the hole-rich domains should maintain the size value $4a$. The same conclusion has been obtained for the incommensurate CDW observed in high $T_c$ when analyzed in direct space\cite{McMillan76,Mesaros16}. This comparison suggests a tight relationship between the size of the domain and the period of a CDW in the 2D metallic state, as observed in the CO state where our data have determined the size $4a$ for the domains and the period $4a$ for the CDW.
Finally, we emphasize that, in this 2D metallic state, the law of $Dq^2$ is connected to the excitations of the hole-poor domains with a "wave" or unlocalized  charge character along the [110] direction and to excitations of bipolarons with a localized character intertwined with the hole-poor domains along the MnO bond directions, reminding of the nodal, anti-nodal dichotomy observed in cuprates at high energy.

The origin of these observations is obtained from our phonon spectrum studies at $T<T_{co}$ and $T>T_{co}$.\\

\begin{figure}[t]
\includegraphics[width=8cm]{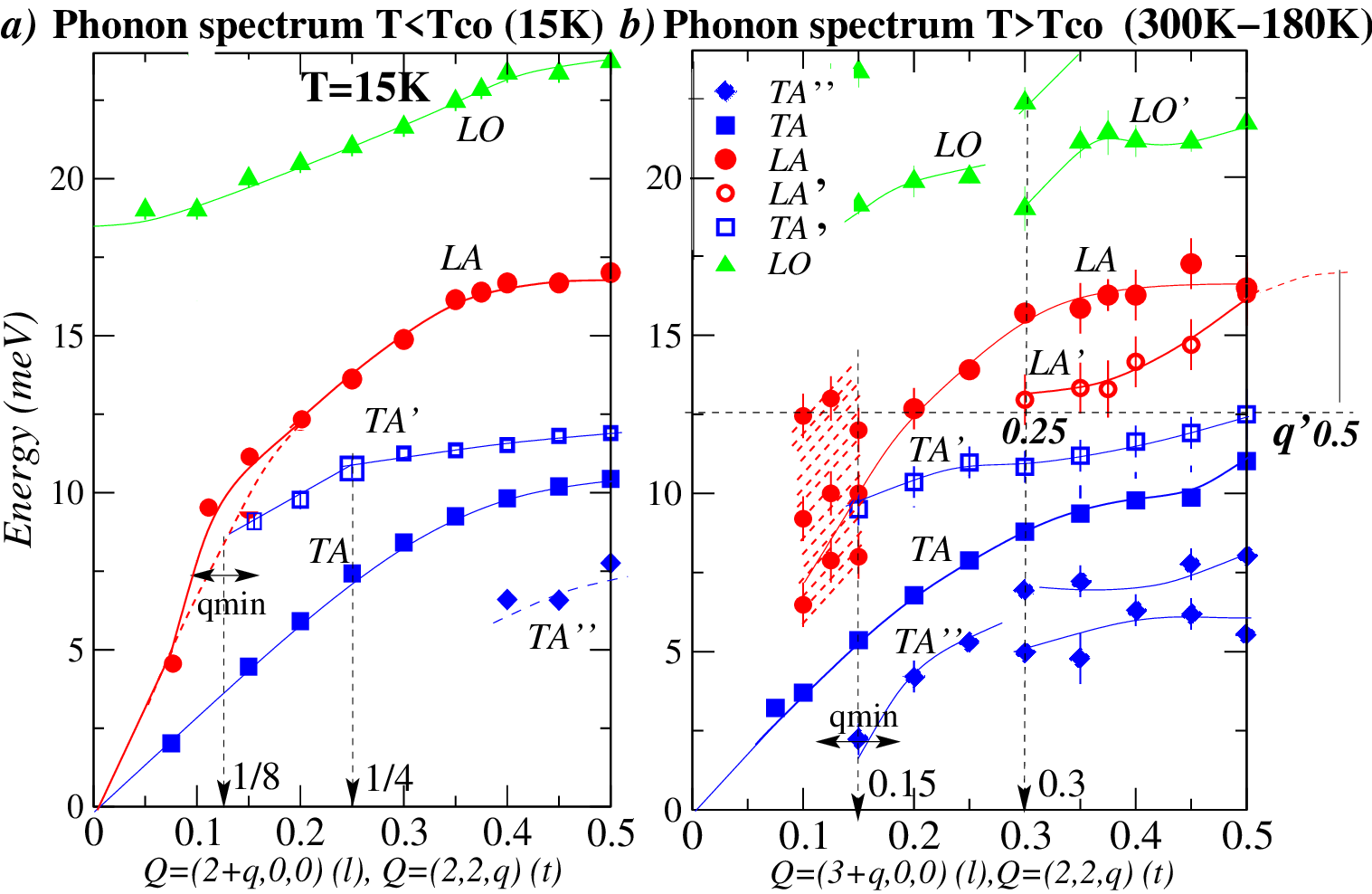}
\caption{\label{figure5} 
Phonon spectra along [100] determined in longitudinal (red and green color) and transverse (blue color) configurations with: in a) at 15K ($T<T_{co}$) one additional branch labeled TA', and, in b), at 300K and 180K ($T>T_{co}$) four additional branches labeled LO' and LA' ($q_{min}\approx 0.3$ rlu), TA' and TA" ($q_{min}\approx 0.15$ rlu). The $q'$ wave-vector is defined by $q'=(0.3/0.25)^{-1}q$ (see the text). The dashed area indicates that the LA(q) branch splits into three branches at $q\le 0.15$ rlu. In a) and b), the continuous and dotted lines are guides to the eye. The dotted vertical lines point to the characteristic values $q_{min}$=0.125 and $q=0.25$ rlu for TA' ($T<T_{co}$) and $q_{min}$=0.15 for TA' and TA" , $q_{min}$=0.3 for LO' and LA' ($T>T_{co}$). The horizontal arrows represent their estimated error.}
\end{figure} 

\begin{figure}[t]
\includegraphics[width=8cm]{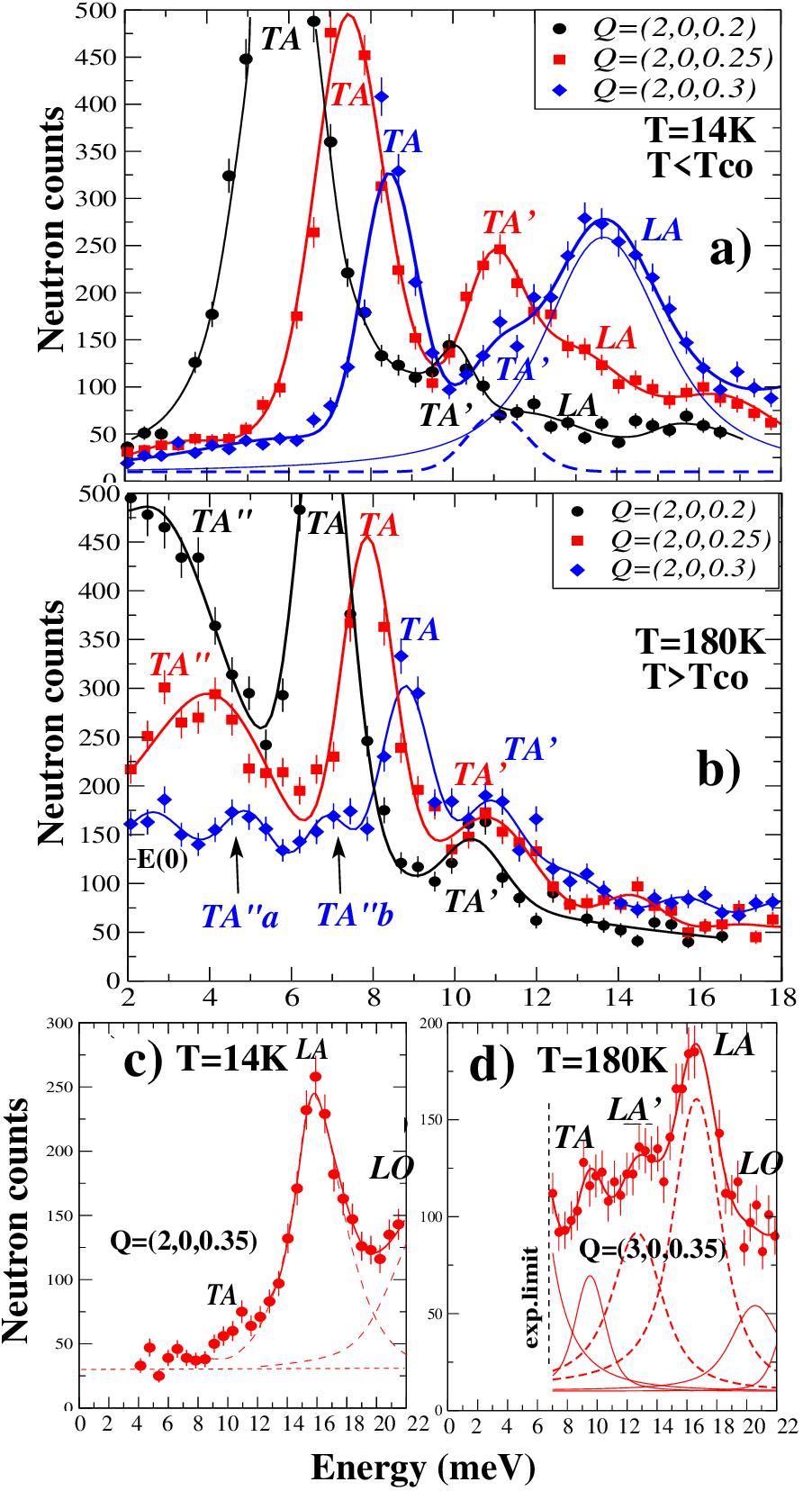}
\caption{\label{figure6} 
a) and b): Comparison of the spectra measured in the transverse configuration at values $q=0.2$, $q=0.25$ and $q=0.3$ at $T<T_{co}$ (14K, upper panel) and $T>T_{co}$ (180K, lower panel, with spectra shifted by 25 counts for clarity). This comparison shows that, at $T<T_{co}$, the TA' branch exhibits a maximum intensity  at $q=0.25$ and that the LA branch acquires a large intensity in the transverse configuration for $q\ge 0.25$. c) and d) Comparison of raw data obtained in longitudinal configuration at $q=0.35$, for $T<T_{co}$ (14K left panel) and $T>T_{co}$ (180K, right panel), showing the occurrence of the LA' mode at $T>T_{co}$. }
\end{figure}

{\bf 2- The additional branch of phonon excitations in the 3D ferromagnetic state ($T<T_{co}$)}

At $T<T_{co}$ the values of the lattice parameter in the pseudocubic structure are very close to each other and the Jahn-Teller effect is nearly suppressed\cite{Argyriou96,Pinsard97,Hennion05}. 

Fig.\ref{figure5}-a shows the existence of an additional phonon branch TA'(q) lying between the TA(q) and LA(q) branches observed in transverse configuration along the [100]+[010]+[001] directions superimposed by twining. This branch TA' gives evidence for the two characteristic wave vectors $q=0.125$ and $q=0.25$. First, this branch disappears at $q_{min}$=0.125 +/-0.025 rlu (see raw data in Ref.\onlinecite{Supp-Mat}, Fig.SM-7 and Fig.SM-8). Near this wave vector the longitudinal branch LA seems to be repulsed from TA' suggesting interaction of these modes with an anti-crossing behavior as shown in Fig.\ref{figure5}-a. 
This value $q_{min}=0.125$ reveals that the cubes $4a$x$4a$x$4a$ observed by magnetic excitations have a charge-phonon coupling origin.  
The wave vector $q=0.25$ rlu is outlined by the highest intensity $I_{max}$ of the branch TA'. This maximum intensity occurs at $T\le T_{co}$ (see the comparison of the raw spectra of the TA' branch in Figs.\ref{figure6}-a and \ref{figure6}-b).
In our previous study at $x=0.2$ (see Introduction), the maximum intensity of the TA* branch observed at $q_{min}=0.25$ has been related to stationary excitations of 1D orbital polarons of size $2a$ in contact along chains. In the same way, in the present 3D case where the TA' branch is nearly q-independent for $q\ge 0.25$ this maximum intensity is attributed to nearly stationary excitations of 3D orbital polarons of size $2a$ in contact along the three MnO directions. 

In conclusion, the lattice excitations determine large cubes of side $4a$ consisting of eight cubes corresponding to orbital polarons of size $2a$. This is the same picture as that obtained by magnetic excitations and observed during their motion. We present on the Fig.\ref{figure4}b the two identical cubes outlining ferromagnetic (Fig.\ref{figure4}b-1) and nuclear (Fig.\ref{figure4}b-2) coupling between polarons. This ordered state of ferromagnetic orbital polarons is responsible for the coincidence of the magnetic and acoustic lattice excitation, which also manifests itself by a new intensity of LA in the transverse configuration for $q\ge 0.25$ (compare the raw data of Fig.\ref{figure6}-a and b, and of Fig.SM-9 in Ref.\onlinecite{Supp-Mat}).  However, because of twinning, the fundamental information on the period $8a$ along the two MnO directions of the planes is missing, which in contrast outlines the powerful probe of the charge correlations provided by the orbital fluctuations (magnetic excitations).

The proposed 3D superstructure unlocks the ordering vectors (1/8 1/8 1/2) and (1/8 1/8 1/4), corresponding to the periodicity of, respectively, magnetic (F)  and lattice (nuclear) coupling. To our knowledge these ordering vectors have not been reported in the literature, probably because of their low intensity as compared to the other observed ordering vectors. This may be a challenge 
for future experiments to confirm the proposed vectors with the (1/8 1/8) in-plane component. 
\\

{\bf 3- The additional phonon branches observed at $T>T_{co}$ in the 2D ferromagnetic and metallic state.}

At $T>T_{co}$ the lattice parameters of the pseudocubic structure exhibit well-distinct values and a large Jahn-Teller effect is observed\cite{Argyriou96,Pinsard97,Hennion05}. 
Fig.\ref{figure5}-b presents evolution of the lattice dynamics above the CO transition. There, three additional branches are observed in the restricted range of values $q$ that give evidence for the two characteristic wave vectors $q=0.15$ and $q=0.3$.
Along the [100]+[010]+[001] directions of $q$, LA (q) behaves differently depending on the range of the wave vector. For $q<q_{min}$=0.15 rlu, the three branches indicate that LA(q) is distinct for the three directions of the MnO bond, superposed by twinning (see the raw data shown for $T>T_{co}$ and $T<T_{co}$ at $q = 0.15$ in Fig.SM-10 of Ref.\onlinecite{Supp-Mat}). For $q>q_{min}$=0.15 rlu, a single branch LA (q) remains that becomes broad at $q=0.25$ rlu. The same wave vector $q_{min}\approx 0.15 +/-0.05$ rlu manifests itself as the point where the two other transverse branches TA' and TA" come out. In fact, due to LA contamination, only the value $q_{min}$ of the branch TA" can be experimentally determined (see the raw data in Fig.SM-11 of Ref.\onlinecite{Supp-Mat}).

Another characteristic wave vector appears at twice this value. The value $q_{min}\approx 0.3$ points to the emergence of the additional LA'(q) branch just below the LA(q) main phonon branch (compare the raw data of Fig.\ref{figure6}-c and \ref{figure6}-d for $q=0.35$. See also raw data at several q values in Fig.SM-12, Ref.\onlinecite{Supp-Mat} and compare with the [110] direction in Fig.SM-13, Ref.\onlinecite{Supp-Mat}). Also, the additional TA" branch apparently splits into two branches at $q\ge 0.3$ (see the raw data of Fig.\ref{figure6}-b  at $Q=(2,0,0.3$).

The data collected above $T_{co}$ point to an "apparent" incommensurate size $a'$ in the 2D metallic regime scaled with respect to the lattice parameter $a$ as $a' = (0.25/0.3)a \approx 0.83a$, with the corresponding zone boundary $q'_{zb}=(0.25/0.3)^{-1}q_{zb}$ of a fictitious Brillouin zone.
With this new zone boundary $q'_{zb}=0.5$ equivalent to $q_{zb}=0.6$, the observed concave dispersion curve of LA'(q) makes sense if extended up to this zone boundary, as shown by the horizontal dotted line in Fig.\ref{figure5}-b. 
It reveals an inflection point in the middle of the [0.25-0.5] range of $q'$ (or the [0.3-0.6] range of $q$) and therefore a new elastic force at the interface between two adjacent domains of size $2a'$.
As for the magnetic excitations where the same value $q_{min}=0.3$ is observed\cite{McMillan76} we attribute this discommensuration to the varying sizes of the hole-poor domains, lattice-locked, intertwined with the hole-rich domains which maintain their size $4a$. 
These observations provide a full characterization of hole-rich bipolarons of size $4a$, resulting from the structural and antiferromagnetic pairings of hole-rich orbital polarons of size $2a$ in a metallic state.

In contrast to LA, the branch TA of the effective medium is still observed at any value of $q$. There, the two additional branches TA'(q) and TA"(q) observed above and below the branch TA with $q_{min}$=0.15 can be attributed to the two expected transverse excitations, "out-of-plane" and "in-plane", of the bipolarons. A splitting of this latter branch occurs on the scale $q \ge 0.3$ of the polaron, showing that the transverse coupling does not play the same role as the longitudinal coupling in the pairing of the polarons. A similar anomaly is observed in the first optical branch reported in Fig.\ref{figure5}-b which corresponds to the vibrations of the cubes of the La/Sr atoms with respect to those of the Mn sites, obtained by translation of the vector $(1/2,1/2,1/2)a$. By the way, this latter anomaly provides the thickness value $a$ of the bipolarons along the $c$ axis. These incommensurate values of $q_{lim}$ being the same as those of the LA' branch, they characterize the same bipolarons of size $4a$ along the two MnO bond directions of the planes.

Actually the best proof of the in-plane $q=(1/8,1/8)$ structure of the bipolarons observed at $T<T_{co}$ appears along the diagonal directions [110] + [101] + [011] of the pseudocubic structure reported in Fig.\ref{figure1}-b. There a strong anomaly is observed between the branches TA and LA, common to the twin domains, at the
value $q=0.125$ outlined by a rectangle in this figure (see raw data in Fig.\ref{figure1}-c). At this value of $q$, the LA and TA excitations exhibit enhanced intensities and shifted energies, corresponding to a tendency to a crossing behavior. This anomaly reveals a periodic modulation of the charge-phonon coupling at $q=(1/8,1/8,0)$ and therefore appears as a precursor state of the in-plane structure of the ordered bipolarons.  

We notice that the spatial characteristics (size, distance) of the bipolarons obtained at $x=1/8$ in the basal planes are the same as those of the ferromagnetic "platelets" observed at $x=0.06$ in a disordered state, static or quenched\cite{Hennion98,Hennion00}. The main difference between the two doping values $x=0.06$ and $x=1/8$ appears along the axis {\bf c} where the thickness value obtained at $x=0.06$ ($7$\r{A} $\approx 2a$) is twice as great as its value, $a$, at $x=1/8$\cite{Hennion00} leading to the common origin of hole-rich bipolarons at the two doping concentrations.

In conclusion of this section, the additional branches of acoustic phonons observed in the range of large values of q ($q_{lim}\ge 0.125$) at $T<T_{co}$ along the directions of the MnO bonds agree with the existence of large cubes $4a$x$4a$x$4a$ resulting from the ordering of orbital polarons $2a$x$2a$x$2a$. Corresponding magnetic excitations that explicitly visualize hole-poor domains reveal the hole-rich character of the orbital polarons ($2a$), their F coupling in forming the domains of size $4a$, and their intertwining with the hole-poor domains along the two MnO bond directions of the planes (period $8a$). They determine a chessboard of hole-rich domains $4a$ x $4a$ in each ($a, b$) plane with alternation of hole-rich and hole-poor domains along the {\bf c} axis. The full 3D charge ordered state is characterized by two families of same cubes of octopolarons of side $4a$, obtained by considering the magnetic excitations also along the diagonals [111]. This superstructure of charges is intertwined with CDW of collective charge origin indicated by gaps in the $Dq^2$ law of the spin waves. 
The bipolaronic origin of the domains $4a$ is provided by the dispersion with $q$ of the additional lattice and magnetic excitations observed in the 2D metallic state. As discussed in layered cuprates concerning the CDW, the apparent incommensurate size of the bipolarons ($0.83$x$4a$ what corresponds to $q_{lim}=0.15$), can be attributed to their intertwining with hole-poor domains of various sizes, lattice locked. 
These observations go beyond the pictures obtained from previous work that describe the charge inhomogeneous state either as a superstructure of polarons\cite{Yamada96}, a transition between two orbital structures\cite{Endoh99}, or a Peierls transition\cite{Wei10}.\\

{\bf III Discussion: A model for high $T_c$ cuprates}.

 The advantage of the magnetic excitations in \la\ for getting the local magnetic correlations, charge-induced, along several symmetry directions is lost in cuprates in which the spins exhibit an antiferromagnetic structure. Moreover, the collective 3D character of these local excitations required for their observations cannot be expected in the layered structure that is necessary to obtain superconductivity. The present observations in \la\ therefore provide a unique opportunity to obtain a realistic model of bipolarons in the CuO planes of cuprates that share the same structure of hybridized orbitals $p-d$ as in the MnO planes.
 
\begin{figure}[t]
\includegraphics[width=8cm]{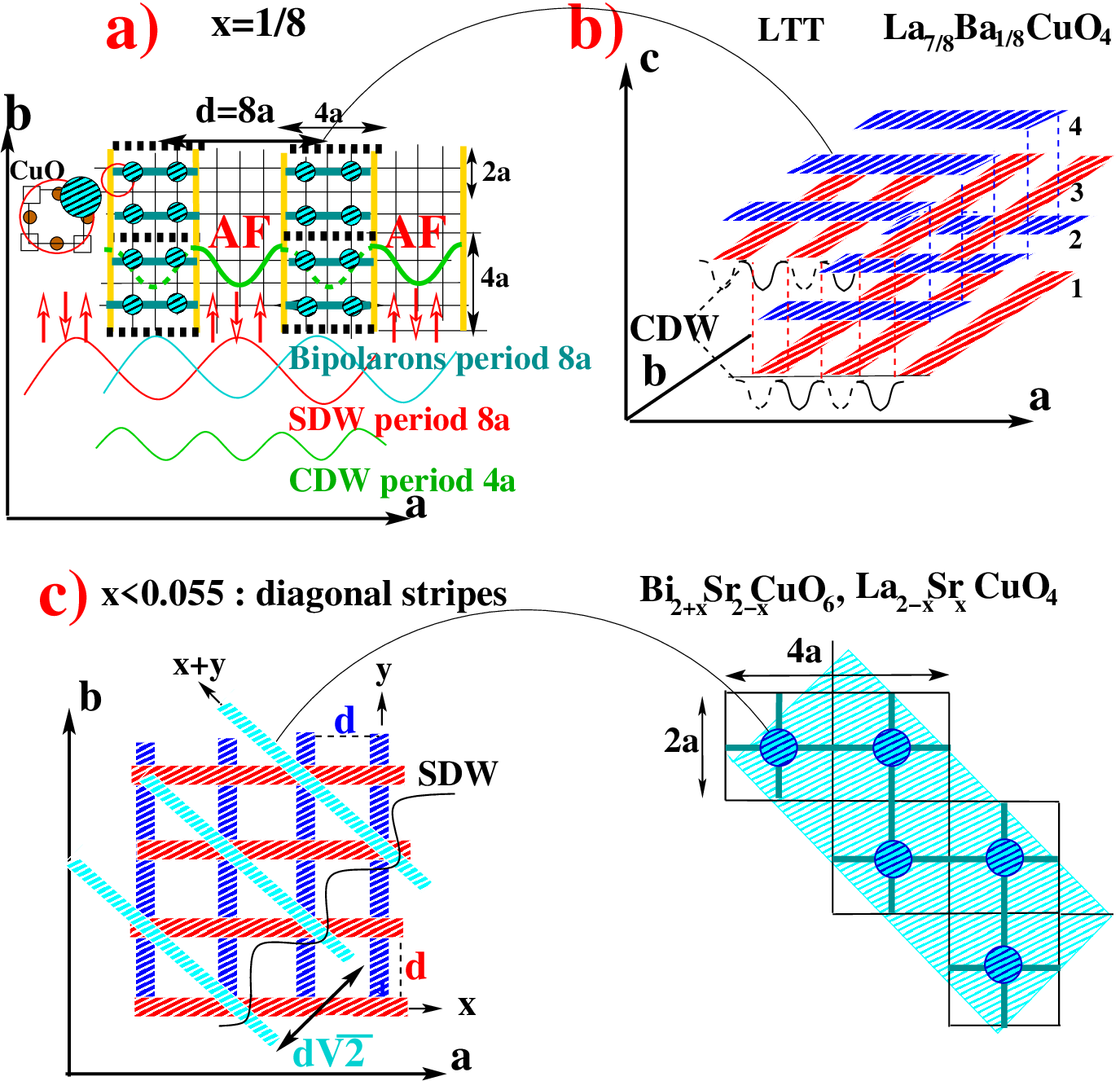}
\caption{\label{figure7} {\bf Model of ordered state of bipolarons in some high $T_c$ cuprates }
{\bf a)} At $x=1/8$: the stripes of hole-rich bipolarons of size $\approx 4a$ are intertwined with stripes of AF spins (hole-poor) of same size $\approx 4a$ along the CuO bond directions. For simplicity, the charge density is sketched by a large circle on the Cu center of the polarons instead of extending on the four surrounding oxygen atoms. The CDW of period $4a$ is sketched by a green line. {\bf b)} Alternation of the hole-rich (bipolarons) and hole-poor (AF spins) stripes every two planes that can be related to a short range period of four layers or $2c$ parameters for the CDW along {\bf c} as observed in La$_{2-x}$Ba$_x$CuO$_4$, from Ref.\onlinecite{Berg07}. {\bf c)} Left hand side: co-existence between the two types of stripes structure of bipolarons, namely along $x$ or $y$, with an alternation along $\bf c$ (blue and red lines as at $x\ge 0.055$), and along the diagonal $x+y$ (cyan lines as at $x\le 0.05$). Right hand side: zig-zag chains of bipolarons are suggested by us for diagonal stripes.}
\end{figure}

Several studies have outlined similarities in the anomaly of the oxygen bond stretching mode observed at $E\approx 60$ meV\cite{Reischardt99,Pintschovius02,Reznik06,Zhu15,Miao18,lin20} and on Fermi surfaces\cite{Yoshida03,Lanzara01,Shen08,Mannella05,Dessau06} in which a strong charge-phonon coupling has been observed and possible local lattice distortions\cite{Egami09}. The direct correspondence between the nodal, anti-nodal dichotomy observed at high energy in cuprates and the present low excitations of bipolarons has been outlined in Section II-I.
In \la, the minimization of the Coulomb energy is obtained by the alternation along the {\bf c} axis between the "hole rich" and "hole poor" domains of the chessboards. 
The same should be true for the biaxial structure of stripes in which the bipolarons should exhibit a linear size $4a$. 
Therefore, testing the present model of bipolarons at $x=1/8$ in the CuO planes implies consideration of a 2D superstructure of stripes of "hole-rich" bipolarons of linear size $4a$ in the plane, distant by $2a$ along the stripe (one orbital polaron) intertwined with stripes of AF-arranged spins, "hole-poor", of same width $4a$ as sketched in Fig.\ref{figure7}-a. The charge density, which is known to lie mainly on oxygen sites, is sketched by large "on-site" circles for simplicity. The size $4a$ of the hole-rich bipolaron that determines the superconducting coherence length $\xi=1.5$ nm is twice that of the Jahn-Teller bipolarons ($2a$) previously considered\cite{Kabanov05,Muller14}. At $x=1/8$, the stripes of the bipolarons should induce a modulation of the AF spin structure that leads to four superstructure peaks of scattering intensity in the wave vector $\delta$(x=1/8)=1/8, along [100] and [010] away from the position of the AF peak $(\pi,\pi)$ or $(1/2,1/2)$ in our notation. As discussed for Jahn-Teller bipolarons\cite{Muller14}, the coherence of the AF spin structure across the stripes of bipolarons could be preserved thanks to the AF pairing (singlet state) of the bipolarons.

In common theories, 2D bipolarons are expected to form a narrow band and, being bosons, to condense into a superconducting state in a small band filling, following the $T_c\propto n_b(T)/m^*$ relation where $m^*$ is the effective mass in the plane\cite{Alexandrov81,Aubry95,Alexandrov96,Proville99,Alexandrov12,Lakhno16,Zhang22}. The density of bipolarons $n_b(T)$ is believed to occur in the pseudo-gap phase and to reach the form $n_b\propto x$ at the $T_c$ value\cite{Alexandrov12} in agreement with the universal law for $T_c$ determined by muon relaxation experiments\cite{Uemura89}. 

The consequences of this model of bipolarons in the dependencies on doping of i) CDW(x), SDW(x), $T_c$(x), ii) on the downward dispersion of the hourglass spectrum and iii) in the extension of this model to low doping are considered successively, specifically for the "214" family, which exhibits the simplest structure.
 
i) In the inhomogeneous spin-charge model sketched in Fig.\ref{figure7}-a, 
the spin density wave (SDW) $\delta$=1/8 results from the intertwining of the AF-arranged spin stripes and the stripes of hole-rich bipolarons with the same width $4a$ where the involved AF spin correlations keep in phase.
In fact, the appearance of a CDW accompanied by lattice distortion is another possible consequence of hole doping with charge-lattice coupling, with a collective charge origin\cite{Alexandrov81}. 
Therefore, the universal "1/8" anomaly could correspond to the peculiar situation that occurs when the size of the hole-poor domain intertwined with the bipolarons becomes equal to the size of the hole-rich bipolarons ($4a$). As in Section II-2 for \la, this situation allows the stabilization of a long-range CDW with $q_{CDW}$ = 1/4 rlu. This modulated charge density expected to be coupled to lattice distortion, intertwined with the localized effect of the charges or bipolarons is another consequences predicted from charge-lattice coupling\cite{Alexandrov81}. It should favor long-time and long-range spatial correlations, leading to a decrease of the bulk value $T_c$\cite{Li07,Kofu09,Chang12,Keimer15,Wen19}. 
 
The most spectacular effect is observed in the LTT La$_{7/8}$Ba$_{1/8}$CuO$_4$\cite{Tranquada95} where the value of $T_c$ measured by the Meissner effect drops to $\approx 4K$ and increases when a disorder is introduced\cite{Modenbaugh88,Li07,Berg07,Berg09,Leroux19}. The observation of a period of four layers (two lattice parameters) along the $\bf c$ axis for the CDW agrees with the existence of hole-rich stripes intertwined with hole-poor stripes in the planes\cite{Berg07}. As sketched in Fig.\ref{figure7}-b, the alternation required to minimize the Coulomb energy along the {\bf c} axis includes every two planes (1+3, 2+4 {\it etc.} as shown in the Fig.\ref{figure7}-b), with parallel and shifted stripes. In this compound, the variation with temperature of $q_{CDW}$ from $q_{CDW}$ = 0.235 rlu, long range, at low temperature to $q_{CDW}$=0.3 rlu, short range, at higher temperature\cite{Miao18}, recalls the temperature variation of the $q_{min}$ value of the lattice and magnetic excitations of bipolarons reported in the present study of \la\ (see Section II-1) outlining the  relationship between the wave vector $q_{CDW}$ interpreted in direct space\cite{Mesaros16} and the scale $4a$ of the bipolarons. 

Due to their intertwining, the ordering of hole-rich bipolarons at $x=1/8$ also optimizes the correlations of the SDW. This is observed in La$_{2-x}$Sr$_x$CuO$_4$ where, at $x=1/8$, a long-range static component of the SDW is observed which coexists with a dynamic short-range component\cite{Kofu09}. A similar effect is obtained by applying a magnetic field\cite{Lake02}. At $x=1/8$, this intertwining yields the relation $\delta=q_{CDW}/2$. The present model of ordered stripes of bipolarons (period $8a$) stabilized by a CDW of period $4a$ may provide a physical origin of the $q=1/8$ rlu charge modulation observed in the vortex halo of Bi2212 cuprate compound by applying a magnetic field\cite{Edkins19} what was interpreted as a pair density wave\cite{Agterberg20}.
The intertwining between the CDW of period $4a$, the SDW of period $8a$ and the presumed order of the PDW with a specific scale $8a$ has also been established in La$_{2-x}$Sr$_x$CuO$_4$ doped with Fe where an SDW coexists with the CDW\cite{Huang21}.

ii) Turning to the hourglass magnetic spectrum, we recall that the magnetic fluctuations of the low energy downward dispersion curve which departs from the (1/2±$\delta$,1/2±$\delta$) magnetic peaks are strongly anisotropic\cite{Hinkov07,Matsuda08,Fujita12}. 
The downward dispersion reveals therefore a new coupling between the AF spins arising from a narrow band of bipolarons in the metallic and superconducting state with a long range character. It competes with the local magnetic coupling of super-exchange origin inside the hole-poor stripes of AF spins with corresponding dispersion curve of energy shifted up at higher energy\cite{Tranquada13} above $E_{cross}$. This long range coupling between spins should be responsible for the disappearance, at $x\approx 0.03$\cite{Wakimoto99}, of the AF peak at ($\pi,\pi$) characteristic of the undoped parent compound.  The magnetic energy gap that opens at the bottom of the downward dispersion curve for $T<T_c$\cite{Kofu09,Tranquada13} is consistent with a coupling between the spin and charge order parameters in the superconducting state. 
  
iii) The present model of stripes of bipolarons can be extended to low doping as follows. The parameter $\delta(x)$ corresponds to the periodic distance d=$a/\delta$ between the stripes of bipolarons. The linear variation of $\delta(x)$ approximately observed in YBa$_2$Cu$_3$O$_{6+x}$ and well obeyed in La$_{2-x}$Sr$_x$CuO$_4$ and Bi$_{2+x}$Sr$_{2-x}$CuO$_6$\cite{Yamada98,Fujita02,Enoki13} therefore corresponds to the decrease of this periodic distance from infinity at $x=0$ to $8a$ at $x=1/8$, as the planes are progressively filled by stripes of bipolarons with doping. Consequently, the band of bipolarons intertwined with hole-poor domains is filled linearly with bipolarons by doping, leading to the two linear relations $\delta(x)=x$ and $T_c \propto \delta$\cite{Yamada98}.

Actually, the relation $\delta(x)= x$ holds even at $x<0.055$ where the SDW rotates along one diagonal direction (with $\delta$ expressed in rlu units) and allows a regime of co-existence of the two types of direction\cite{Fujita02,Enoki13}.
In the present model, the rotation corresponds to a decrease in $\delta$ by the factor 1/$\sqrt 2$ in \r{A} units. It corresponds to an increase of the periodic distance of the stripes of bipolarons by the factor $\sqrt 2$ so that bulk metallic and superconducting properties disappear. The coexistence of two regimes for the stripes, straight (along the CuO bond) and diagonal, corresponds to the situation sketched in Fig.\ref{figure7}-c, left-hand side, where the stripes of bipolarons that run along $(x+y)$ are nucleated at the crossing points between the stripes that run along $x$ and along $y$. 
The continuity of the observations shown by the ARPES experiments along the nodal and antinodal directions\cite{Yoshida03} has led to the proposal of a "staircase" for the CDW\cite{Grannath04}. In the present picture, the continuity suggests rather a zigzag chain of bipolarons (see the sketch in Fig.\ref{figure7}-c) where one charge is shared by two bipolarons as in the chessboard of La$_{7/8}$Sr$_{1/8}$MnO$_3$. In the limit of the large concentrations, the change in behavior observed at $x\approx 0.16-0.17$ in cuprates\cite{Frachet20} as in \lasr means that the present hole-rich, hole-poor model is no longer valid. This conclusion agrees with the existence of a minimal distance of approach between bipolarons as previously observed in a Ca-doped manganite with $x=0.08$\cite{Hennion98} .
\\

In summary, the local excitations of the lattice and magnetic origins observed in pseudocubic \la\ reveal that the structural, magnetic and electric transition observed in $T=T_{co}$ can be interpreted as a transition from hole-rich bipolarons, AF paired, with a size $4a$ intertwined with hole-poor domains of varying size at T>$T_{co}$ towards a 3D charge-ordered state. In the planes, this later consists of a chessboard of F-coupled hole-rich bipolarons $4a$x$4a$ intertwined with hole-poor domains of the same size, leading to a period $8a$ along the two MnO bond directions of the planes. This chessboard of bipolarons is intertwined with CDW arising from a collective charge effect detected by gaps in the $Dq^2$ laws.   
We believe that the picture of the 2D metallic state is valid for the metallic CuO planes of cuprates that share the same hybridized $p-d$ orbital structure. The layered structure implies replacing the chessboard of bipolaron by stripes of bipolarons, so that one gets stripes of hole-rich bipolarons intertwinned with AF spin stripes, hole-poor, rotated by 90 degrees from plane to plane along the {\bf c} axis. This hole-rich, hole-poor model which results from the competition between  attractive and repulsive forces on the charges makes possible a large charge-phonon coupling, necessary to obtain a high critical temperature. As in manganites, the universal "1/8" anomaly should correspond to the peculiar situation where the hole-poor domains intertwined with the hole-rich bipolarons reach the size $\approx 4a$ of the bipolarons. In this way, their fluctuations are reduced by the stabilization of a long-range CDW with wave vector $q_{CDW}=1/4$ rlu that coexists with bipolarons of size $4a$, leading to their ordering with a period $8a$ revealed by a spin density wave of the same period. This model, consistent with the universal law for $T_c$\cite{Uemura89}, can be extended to low doping, leading to Yamada's law $\delta(x)=x$ and $T_c\propto\delta$ observed in La$_{2-x}$Sr$_x$CuO$_4$. 
 Our model of bipolarons advanced in the present paper for the 214 family with ordered stripes should be also valid for highly disordered cuprates such as Bi-based compound Bi2212\cite{Mesaros16}. The existence of a $4a$ scale in the whole pseudo-gap phase of this Bi-based cuprate with a filling process observed by direct imaging\cite{Mesaros16} reinforces our confidence in the reliability and general significance of our model.

Although the origin of the pairing is still debated in cuprates, there exists an approach that the pairing occurs in direct space and that the bosons fill the space by doping. The magnetic fluctuations pairing mechanism of superconductivity in cuprates has been long considered as the most probable contender. Nevertheless up to now this mechanism has not been unanimously and undoubtedly approved as the only valid candidate.
Our approach for lattice-mediated pairing of pre-formed charge objects (bipolarons) does not exclude other mechanisms of superconducting pairing such as spin-fluctuation-mediated interactions but may play its role in one row with the others.
\\

After finishing this manuscript, we have learned of a direct imaging experiment at very low doping in a Bismuth-based cuprate family\cite{Li23}. The authors show the existence of $4a$x$4a$ plaquettes the number of which increases with doping.
However, they also reveal the existence of an internal anisotropic texture of charge density on oxygen sites inside the large plaquette $4a$x$4a$ and its role for setting the direction of the stripe. Their observation means that the stripes consist of $4a$x$4a$ domains aligned in a row (see dashed lines in Fig.\ref{figure7}-a) in place of anisotropic domains $4a$x$2a$ depicted in the same figure. This outlines the role of the size $4a$ characteristic of the present bipolaronic model and the periodicity $8a$ inherent to doping x=1/8. 
As commented by Zaanen\cite{Zaanen23}, we are faced with complex effective pairs, that certainly cannot be fully resolved by a unique scale of superconducting pairing but likely by two distinct scales.  
It cannot be excluded that the $4a$ scale determined by copper sites (plaquettes) and a $2a$ scale represented by oxygen sites compose a tentative platform for realisation of this approach.
\\

 {\bf Methods}
 
  {\bf Sample preparation}
  
 A single crystal of \la\ was grown by the ICMMO "Institut de Chimie moleculaire et des materiaux d'Orsay" at Orsay University (France). The single crystal, twinned, used for inelastic neutron experiments is a cylinder with 40 mm of height and 4 mm of diameter. Complementary measurements of ac susceptibility, resistivity, lattice parameters and oxygen positions as a function of temperature have been previously reported in Ref.\onlinecite{Hennion05}. At $T=T_{co}$, the structural phase transition evolves from a monoclinic structure ($T>T_{co}$) to a triclinic structure ($T<T_{co}$), both structures being close to the orthorhombic one. 
 \\

 {\bf Inelastic neutron scattering experiments}
 
The excitations in \lasr\ at $x=1/8$ were measured by inelastic neutron scattering at the three-axis neutron spectrometers (TAS) 4F installed at the cold neutron source and 2T at the thermal neutron source of the Laboratoire Léon Brillouin (Orphée reactor, Centre d'Etudes de Saclay, France) and the thermal neutron TAS-IN8 (Thermes) at the Institut Laue-Langevin (Grenoble, France)\cite{Piovano23}. The spectrometers largely used open geometry with focussing pyrolytic graphite analysers and monochromators (reflection PG002 at 4F, 2T, IN8) and a silicon monochromator (reflection Si111 at IN8). Depending on the studied energy range, filters of higher monochromatic harmonics were installed in the scattered beam (polycrystalline Be-filter at 4F, oriented PG filters at 2T, IN8).\\

{\bf Ackowledgements}

The authors are very indebted to S. Aubry and L. Proville for enlightening and encouraging discussions, and to F. Moussa for many years of friendly and fruitful collaboration. The authors are grateful to S. Petit for his instructive comments supported by calculations of the neutron-scattered intensity and his critical reading of the manuscript, and D. Bounoua, P. Bourges, and Y. Sidis for stimulating criticisms.
\\

{\bf Author contributions}

M.H. conceived the idea and designed the experiments. Inelastic neutron scattering measurements were performed and the data analysed by M.H., A.I. and B.H. Theoretical ideas and concepts were argued and debated by M.H., A.I. and C.L. The manuscript was written by M.H. with the assistance of A.I. and C.L. and all authors discussed the results and commented on the manuscript.
\\

{\bf Competing interests}

The authors declare no competing interests. 
\\

{\bf Data availability}

 All relevant data are available from the authors upon reasonable request.
 Correspondence and materials requests should be directed to Martine Hennion (martine.hennion02@gmail.com) and Alexandre Ivanov (aivanov@ill.fr). 
 \\

{\bf Additional information}

The "Supplementary Material" is added at the end of the manuscript, after the bibliography list of the main paper.
\\

\newpage
\appendix

\author{M. Hennion}
\affiliation{Laboratoire L\'eon Brillouin, Universit\'e Paris-Saclay, CNRS, CEA, F-91191 Gif-sur-Yvette, France}
\author{A. Ivanov}
\affiliation{Institut Laue-Langevin, 71 avenue des Martyrs, CS 20156 F-38042 Grenoble, France}
\author{C. Lacroix}
\affiliation{Institut Néel Grenoble-Alpes CNRS, 25 avenue des Martyrs BP 166  F-38042 Grenoble, France}
\author{B. Hennion}
\affiliation{Laboratoire L\'eon Brillouin, Universit\'e Paris-Saclay, CNRS, CEA, F-91191 Gif-sur-Yvette, France}

\begin{center}

\end{center}

{\bf Supplemental information to the paper:}\\

{\bf From chessboard of bipolarons of size $4a$ in cubic \la\ to 
stripes of the same bipolarons in layered high $T_c$ cuprates}\\

{\bf I-1 Introduction: Comparison of phase diagrams of La$_{1-x}$Sr$_{x}$MnO$_3$ and La$_{2-x}$Sr$_{x}$CuO$_4$}

\begin{figure}[H]
\includegraphics[width=8cm]{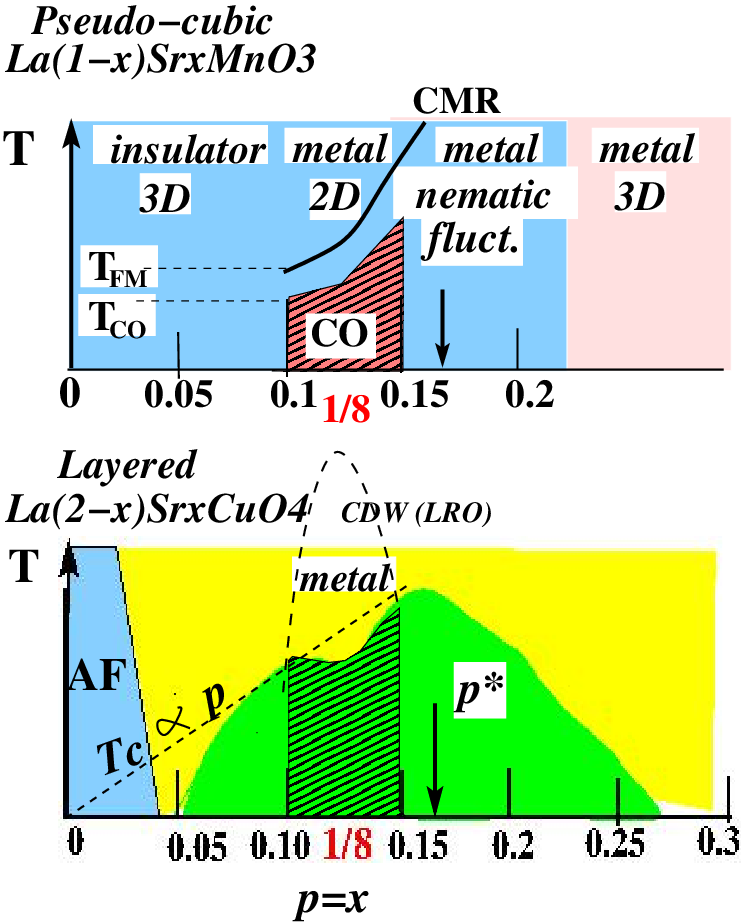}
Fig.SM-1: In {\bf upper figure} is sketched the phase diagram of \la\ in the doping range $x<0.3$ adapted from Ref.\onlinecite{Paraskelvopoulos00} and Ref.86.
Three regimes are observed i) a "canted state" ($0.05<x<0.1$) where "platelets" of size $\approx 4a$, thickness $2a$, with a minimal distance of approach  have been observed (see Ref.86 and Ref.\onlinecite{Hennion98,Hennion00}) ii) a charge ordered state ($0.1<x<0.16$) indicated by the occurrence of superstructures at $q=0.25$ rlu and at $q=0.125$ rlu along $c$ at T<$T_{co}$ (see Ref. \onlinecite{Yamada96} and Ref.87). iii) a regime with optimal magneto-resistance ($0.16<x<0.2$) where interacting chains of orbital polarons have been observed (see Ref.88 and Ref.\onlinecite{Hennion19}.
In {\bf lower figure} is sketched the phase diagram of \lacuo\ adapted from Ref.\onlinecite{Kofu09} and Ref.\onlinecite{Wen19}, showing long-range CDW with a flattening or minimum of $T_c$ in the $0.1<x<0.14$ doping range and a $T_c$(x) $\propto x$ variation up to $x\approx 0.14$. The $x^*=p^*\approx 0.17$ value indicates the transition towards a new regime.
\end{figure}

Fig.SM-1 displays the schematic phase diagram of \lasr\ (upper panel) and of \lacuo\ (lower panel) up to $x\approx 0.3$ adapted respectively from Ref.\onlinecite{Paraskelvopoulos00}, and Ref.\onlinecite{Kofu09} or Ref.\onlinecite{Wen19} . Whereas the two compounds differ at low and large doping where the spin dynamics of \lasr\ exhibits a 3D character, a strong similarity appears in the $0.1<x<0.15$ around x=1/8 where the metallic state of \lasr\ observed at high temperature is 2D\cite{Hennion06}. The hatched areas correspond, for \lasr (upper panel) to a charge-ordered state, ferromagnetic and insulating, and for \lacuo (lower panel), to a concentration range where a CDW with long-range order is observed with a decrease of $T_c$. \\\\

{\bf I-2 Lattice and magnetic excitations observed at $x=0.2$ }

\begin{figure}[H]
\includegraphics[width=8cm]{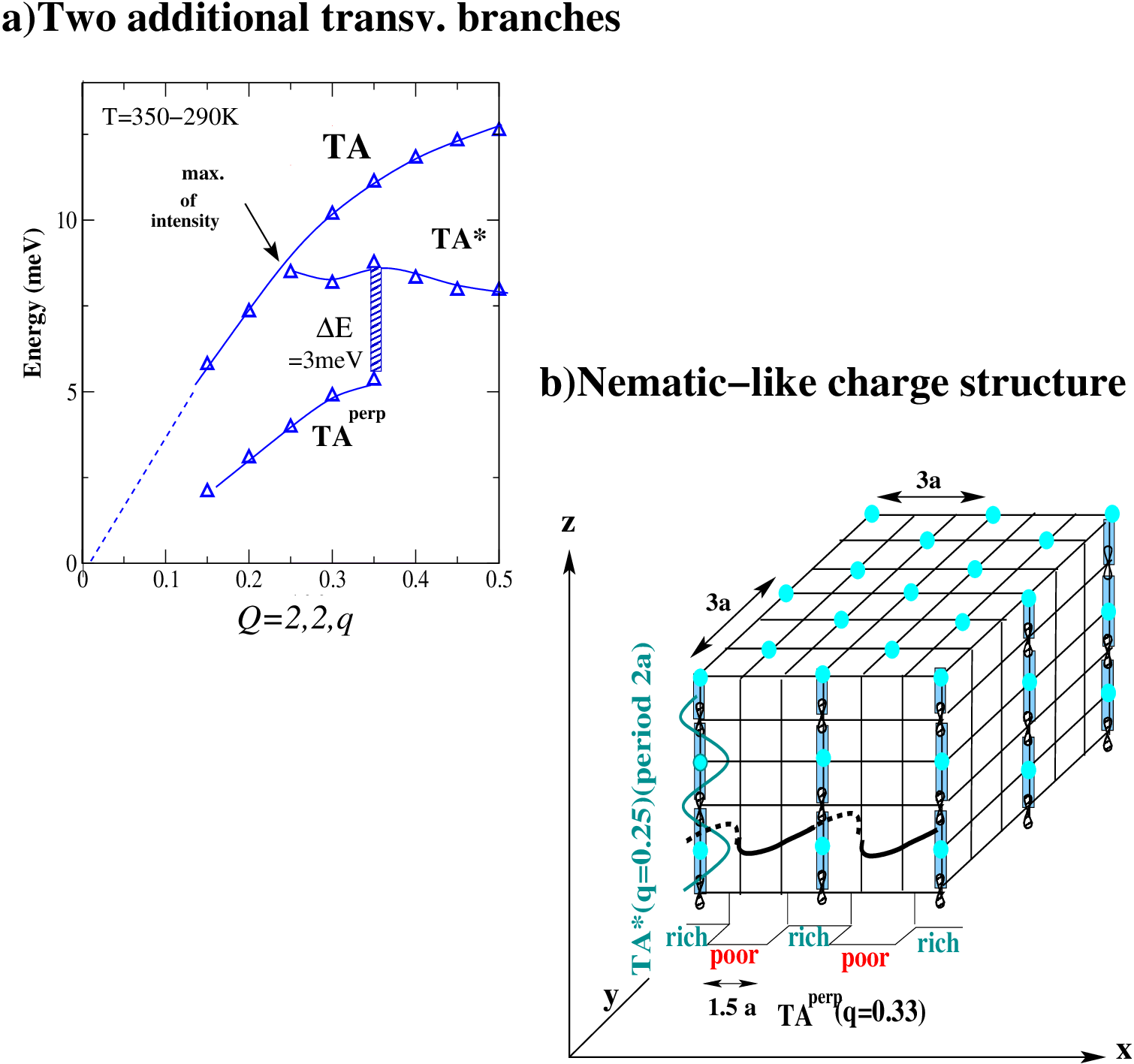}
Fig.SM-2: {\bf a)} Additional transverse acoustic branches $TA^*$ and $TA^{perp}$ observed in restricted ranges of values of q below the main branch TA (Ref.\onlinecite{Hennion19}). They correspond to two perpendicular wave vectors superposed due to the twinning. {\bf b)} Nematic-like structure determined by the limits of the  $TA^*$ and $TA^{perp}$ branches. The stationary branch with the limit $q_{min}=0.25$ ($TA^*$) corresponds to vibrations of domains of size $2a$ in contact. They define chains of 1D orbital polarons along one of the three possible MnO bond directions, arbitrary labeled $z$. 
\end{figure}

\begin{figure}[H]
\includegraphics[width=8cm]{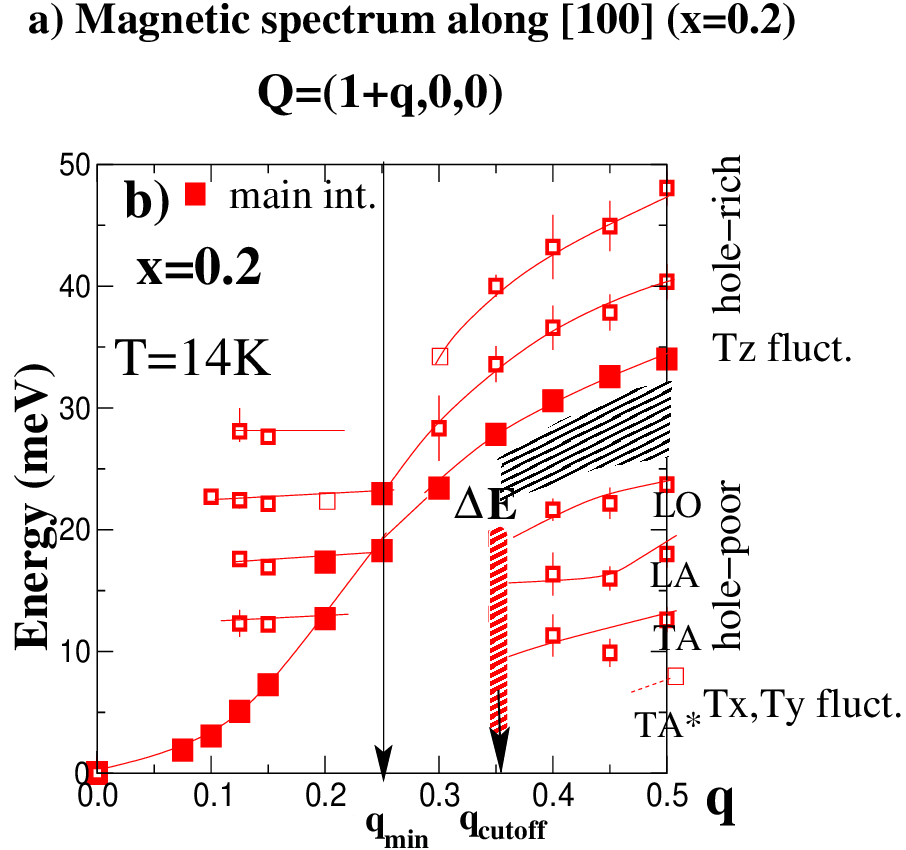}
Fig.SM-3: Magnetic spectrum observed along the MnO bond
directions showing two sets of (q, E) branches in large range of q values with $q_{min}=0.25$ (E>27 meV) and $q_{cutoff}=q_{min}=0.35$ (E<27 meV). The discrete energy spectrum exhibits the two same characteristic q values of the additional acoustic phonon branches and therefore corresponds to the fluctuations of the same superstructure of charges defined by hole-rich domains (1D orbital polarons) observed through $T^z$ fluctuations and hole-poor domains in the direction perpendicular to the chains observed through $T^x$, $T^y$ fluctuations.  
\end{figure}

\begin{figure}[H]
\includegraphics[width=4cm]{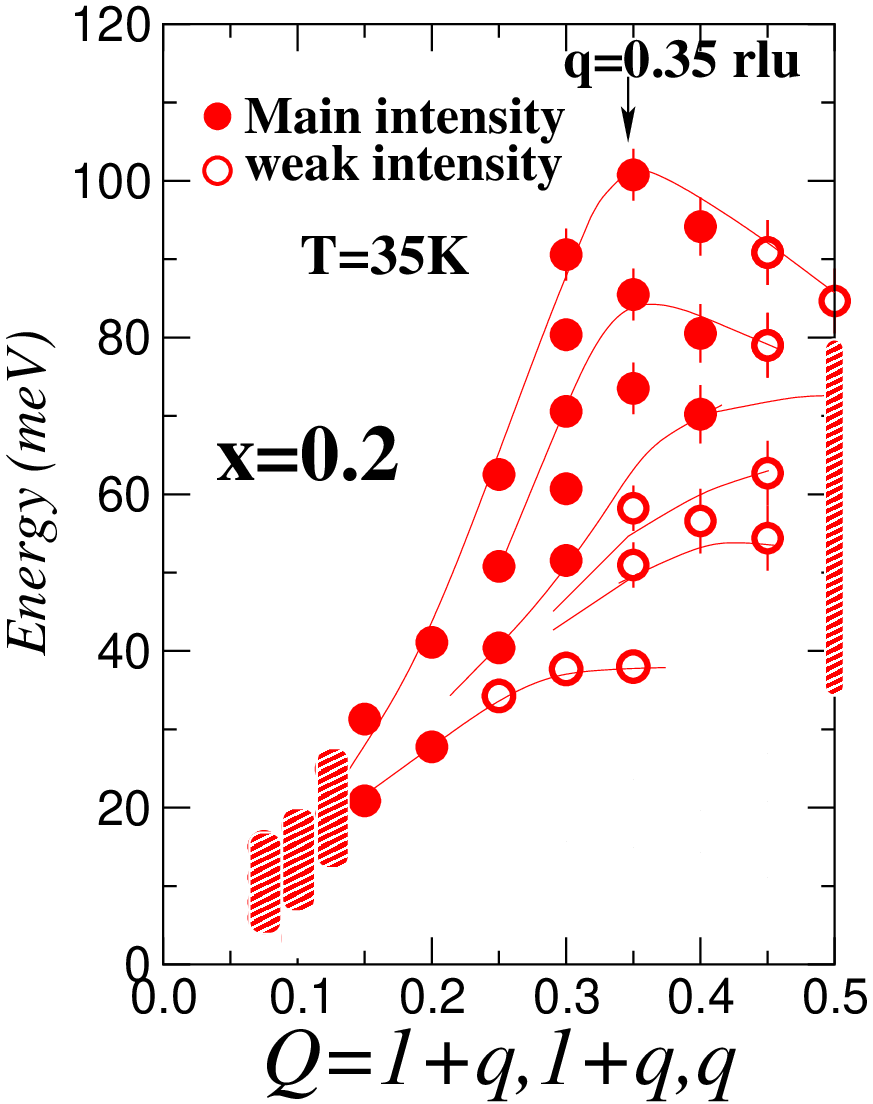}

Fig.SM-4:  Magnetic spectrum obtained along the [111] direction. No regime corresponding to a $Dq^2$ law can be identified. The $q=0.35$ rlu value where the energy decreases, corresponds to the zone boundary of the excitations characteristic of the MnO bond directions (see the text).
\end{figure}

The phonon spectrum observed along the MnO bond direction has revealed a longitudinal acoustic phonon branch $LA^*$ and two transverse acoustic phonon branches $TA^*$ and TA$^{perp}$ in a restricted range of q values, in addition to the LA and TA branches of the pseudocubic structure with high intensity (see Fig.SM-2a, left panel, adapted from Ref.\onlinecite{Hennion19}). They define a superstructure of parallel chains of 1D polarons as follows. The limit $q_{min}=0.25$ rlu of the branch $TA^*$, stationary, with a maximum intensity at $q=0.25$, determines the size $2a$ of orbital polarons in contact along chains so that the excitations of each polaron are in phase. They are arbitrary aligned along $z$ in Fig.SM-2a, right panel. The cutoff $q\approx 0.35$ of the TA$^{perp}$ attributed to the perpendicular wave vector superposed due to twining may be seen as the zone boundary for a fictitious Brillouin zone defined by chains with periodic distance $3a$ along the two MnO directions of the planes perpendicular to the chain directions. At this value, the transverse excitation corresponds to the vibration of the hole-rich and hole-poor domains in phase opposition, each with a size $\approx 1.5 a$. The energy $\Delta$=3 meV corresponds to the energy for aligning the chains or for changing their direction.

The corresponding magnetic excitation spectrum is reported in Fig.SM-3\cite{Hennion06,Hennion19}. 
 
 Beyond the law $Dq^2$, the magnetic spectrum is divided into two energy ranges showing the same characteristic values of $q$ as the two additional transverse acoustic phonon branches $TA^*$ and TA$^{perp}$, and separated by the same energy $\Delta E$=3 meV at the cutoff wave-vector $q\approx 0.35$. The three branches observed above this value correspond to the $T^z$ fluctuations related to the three possible directions of the chain. The four branches observed below this value correspond to the $T^x$,$T^y$ fluctuations of the hole-poor domains intertwined with the chains in the fluctuations perpendicular to the chain direction. This attribution agrees with the variation with the temperature of energy measured at q=0.5 below $T_{FM}$, much sharper for the hole-rich domains than for the hole-poor ones\cite{Hennion19}. 
 
The 1D or uniaxial metallic behavior of the compound at $x=0.2$ is deduced from the observation that the law $Dq^2$ reported in Fig.SM-3, is connected at $q=0.25$ to the upper energy range corresponding to the hole rich domains (chains of 1D orbital polarons), beyond the energy range of the hole-poor ones. This uniaxial character for the charge correlations should be very sensitive to the application of a magnetic field at $T_{FM}$ and therefore can explain the optimum magnetoresistance observed in \lasr   at the concentration $x\approx 0.16$ (Ref.89).

The magnetic spectrum along [111] reported in Fig.SM-4 differs strongly from the spectrum observed along [100] + [010] + [001] reported in Fig.SM-3. Instead of a quadratic law $Dq^2$ in the small range of q values, a multitude of peaks appear, which can be interpreted as a linear combination of excitations observed along the three directions of the MnO bond. In particular, the maximum energy observed at $q=0.35$ rlu corresponds to the zone boundary $q=0.5$ of the MnO bond directions.   
We conclude that the charges do not move along [111] but only along one of the three MnO bond directions of the chains.\\\\

{\bf II Experiments at x=1/8}\\
 
A difficulty inherent to the use of a large single crystal with a pseudocubic structure is the existence of twinned domains, since they lead to the superposition of elastic features as of inelastic ones of non-equivalent symmetry directions. For the elastic peaks observed at $T<T_{co}$, the orthorhombicity is large enough to unambiguously attribute the superstructure peaks observed at q=0.125 rlu and q=0.25 rlu in Ref.\onlinecite{Hennion05} to the {\bf c} direction. For inelastic features of lattice origin, the twinning prevents us from distinguishing between the two MnO bond directions of the planes and those perpendicular to the planes.  This distinction is obtained unambiguously by magnetic excitations in which the MnO directions of the planes ($T^x$,$T^y$ fluctuations) can be distinguished from the MnO direction perpendicular to the plane ($Tz$ fluctuations). \\\\

 {\bf II-1 Magnetic excitations at x=1/8: raw data}\\

All reported spectra correspond to constant $Q$=$q$+$\tau$ scans with varying energy. There, $\tau$ defines the Brillouin zone of the measurement. The major part of the magnetic excitations has previously been reported. They have been measured in the Brillouin zone $\tau$ = (1,0,0), considering that the intensity is proportional to the squared form factor of Mn which decreases with Q. The phonon branches are determined in the zone $\tau$ = (2,0,0) and $\tau$ = (3,0,0) in the transverse or longitudinal configuration in order to optimize the geometrical factor $(Q$x$e)^2$ entering in the intensity (e is the polarisation vector of the excitation). The attribution of the excitations to the lattice excitations can be done unambiguously by considering the temperature variation and the linear dispersion of the branches E(q) observed in the small range of q values. The magnetic character of the excitations measured in the Brillouin zone $\tau$ = (1,0,0) that coincide with the phonon energy values for $q\ge 0.25$ rlu has been verified by polarized neutrons\cite{Hennion06}. The continuous and dashed lines are calculated curves that account for the energy resolution of the spectrometer.
For the excitations of lattice origin, most of the data were obtained during the same experiment with the same experimental conditions. This allows us to use the same background for the spectra (temperature independent) obtained in longitudinal and transverse configurations, if determined at equal counting times.\\

\begin{figure}[H]
\includegraphics[width=8cm, scale=1.2]{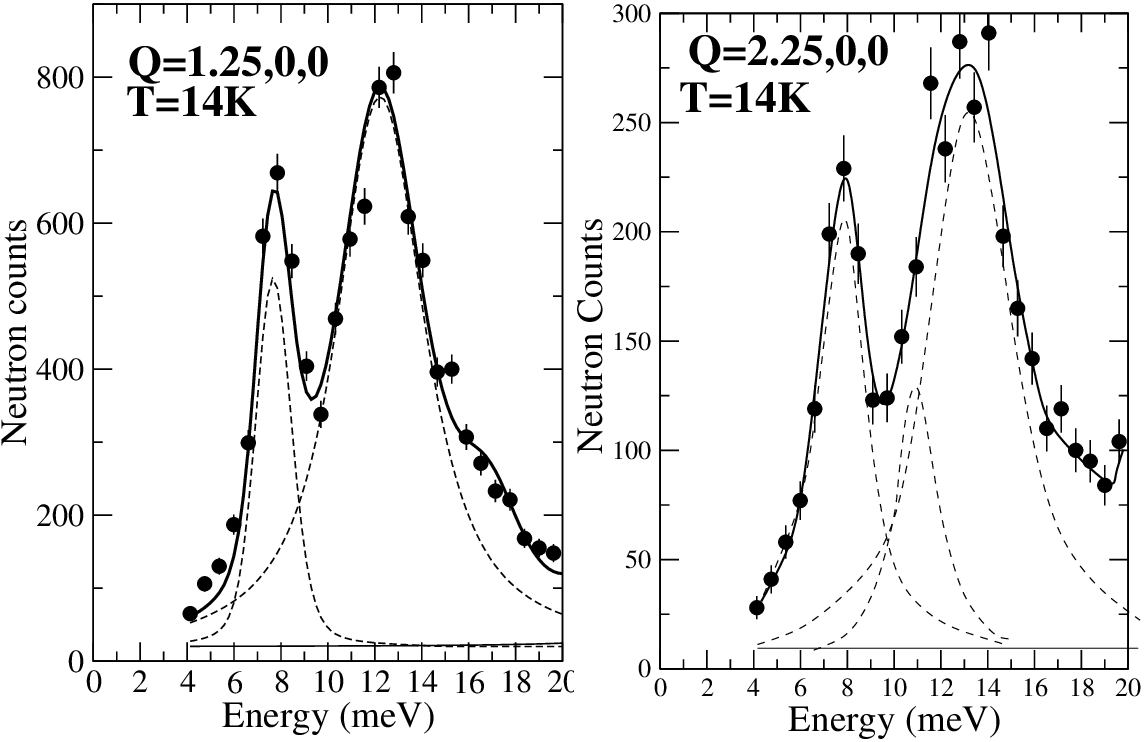}
Fig.SM-5: Comparison between the spectra obtained at $Q=(1.25,0,0)$ (left hand side) and $Q=(2.25,0,0)$ at T=14K ($T<T_{co}$).
\end{figure}

Fig.SM-5 illustrates the coincidence in the values (q, E) between the two excitation peaks of the magnetic spectrum (left panel) measured at Q=(1+q,0,0) and the two excitation peaks corresponding to the TA and LA 
acoustic branches determined in the longitudinal configuration at Q=(2+q,0,0) (right panel). This spectrum has also been measured by the polarized neutrons in Ref. \onlinecite{Hennion06}. At this value $q=0.25$,
there is a large change in the intensity of the magnetic excitations from the lower values of q ($q<0.25$) towards the upper values of q ($q>0.25$) as 
reported in Fig.4-c of Ref.\onlinecite{Moussa03}. This value is used to define the value the $q_{lim}$ value of the hole-rich domains (see the text).\\

\begin{figure}[H]
\includegraphics[width=8.5cm]{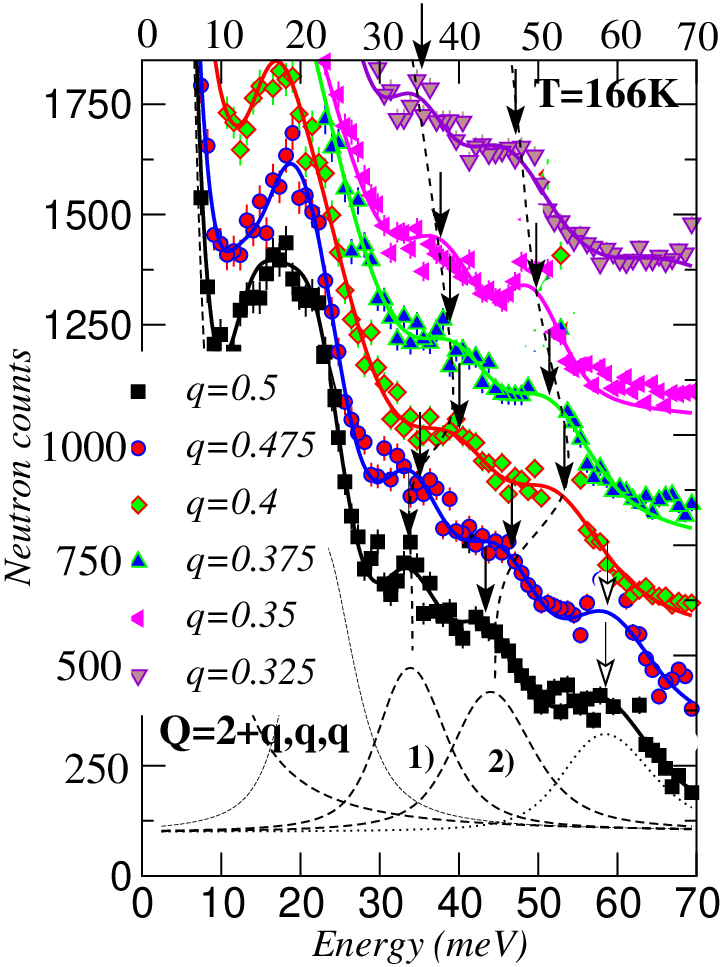}
Fig.SM-6:  Raw data measured along the [111] q-direction. We have checked that these spectra are identical to those measured along [110].  The spectra of each $q$ value have been shifted by 250 neutron counts for clarity. The continuous lines correspond to calculated lines and the dashed lines are guides for the eyes. The full arrows point out the maximums of intensity in the two E(q) dispersed curves.
\end{figure}

Fig.SM-6 displays raw data corresponding to the magnetic excitations along [110] but measured along [111] to open the available energy window. Due to the 2D character of the metallic state deduced from the stiffness constant $D$ measured along the tree symmetry directions, the two directions [111] and [110] are equivalent. In both cases, one measures the excitations arising from the $T^x$ and $T^y$ orbital excitations.
The two peaks observed around E=60 meV at q=0.475 rlu and q=0.5 rlu indicated by the empty arrows are attributed to spurious effects. \\\\
\newline

 {\bf II-2 Additional phonon branch TA' observed at $T<T_{co}$ along the MnO bond direction }\\

\begin{figure}[H]
\includegraphics[width=7cm]{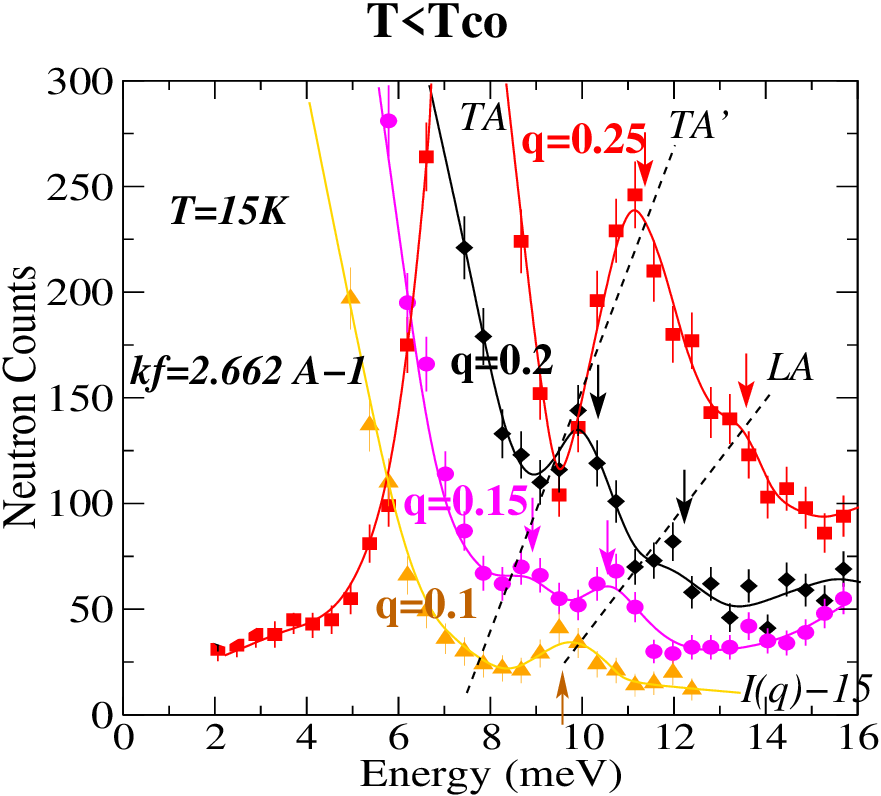}

Fig.SM-7: Raw data obtained in the transverse configuration at the Q=(2,0,q) wave vector values with $q$ = $0.25$,$0.2$,$0.15$,$0.1$ rlu, T=15K ($T<T_{co}$). In this figure, the neutron counts of the peaks corresponding to the main TA branch, not observed on this energy scale, are $\approx 500, 700$ and $800$ for $q=0.25$, $0.2$, and $0.15$ rlu, respectively. The continuous curves are calculated curves using the resolution function of the spectrometer. The spectrum obtained at $q=0.1$ has been shifted by $-15$ neutron counts for clarity. The two black dotted lines outline the dependence with q of the excitations which determine the TA'(q) (upper energy) and LA(q) (lower energy) branches. The TA'(q) branch no longer exists at $q=0.1$ rlu, whereas the LA(q) branch can be measured down to q=0.075 rlu, reported in Fig SM-6. It corresponds to our experimental limit.

\end{figure}

\begin{figure}[H]
\includegraphics[width=4cm]{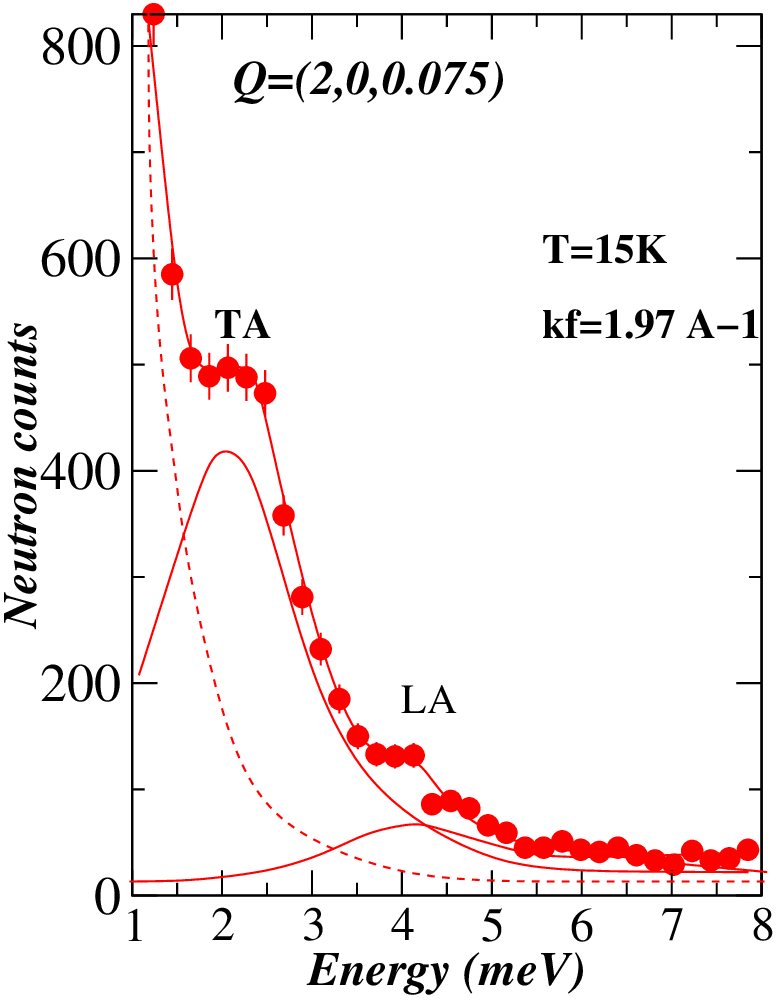}

Fig.SM-8: Raw data obtained in transverse configuration at the Q=(2,0,0.075) wave vector values with constant final wave-vector $k_f=1.97$ $A^{-1}$, T=15K ($T<T_{co}$). The continuous curves are calculated curves using the resolution function of the spectrometer. Only two modes corresponding to the TA and LA branches can be detected.
\end{figure}

The possibility of determining the $q_{min}$ value of the TA' branch is related to the possibility of distinguishing a mode with an energy value intermediate between the TA mode and the contamination of the LA mode. The common existence of these three modes - TA, TA' and LA - is unambiguous not only at $q=0.3$ where the TA' mode appears as a shoulder on the lower energy side of the LA peak (see the spectrum with blue diamond character in Fig.6 of the main paper) but also at $q\ge 0.35$ and beyond this value of $q$ where the LA and TA modes exhibit higher intensity (see Fig.SM-9 at $q=0.35$). At the values $q = 0.25$ and $q = 0.2$, reported in Fig.SM-7, the separation of TA' from the contamination of the LA branch seems more ambiguous. However, the two modes are unambiguously observed at $q=0.15$ both with weak intensity. Only one mode persists at $q = 0.1$. Its attribution to LA contamination can be done by considering the overall evolution of the energy peaks with q indicated by the two dotted lines that link the two arrows defined at each q value in Fig.SM-7. 
From this evolution, we conclude that the branch TA (q) disappears at a value of q between $q=0.15$ and $q=0.1$, while the branch LA can be measured down to $q=0.075$ by using the value $k_f=1.97$ $A^{-1}$ for the final neutron wave vector (our experimental limit, Fig.SM-8). A slight anti-crossing behavior is suggested between LA(q) and TA'(q). We emphasize that the disappearance of TA'(q) between $q=0.1$ and $0.15$ cannot be related to an increase in background, since the latter, mainly defined by the tail of TA(q), decreases with decreasing q.\\

\begin{figure}[H]
\includegraphics[width=8cm, scale=1.2]{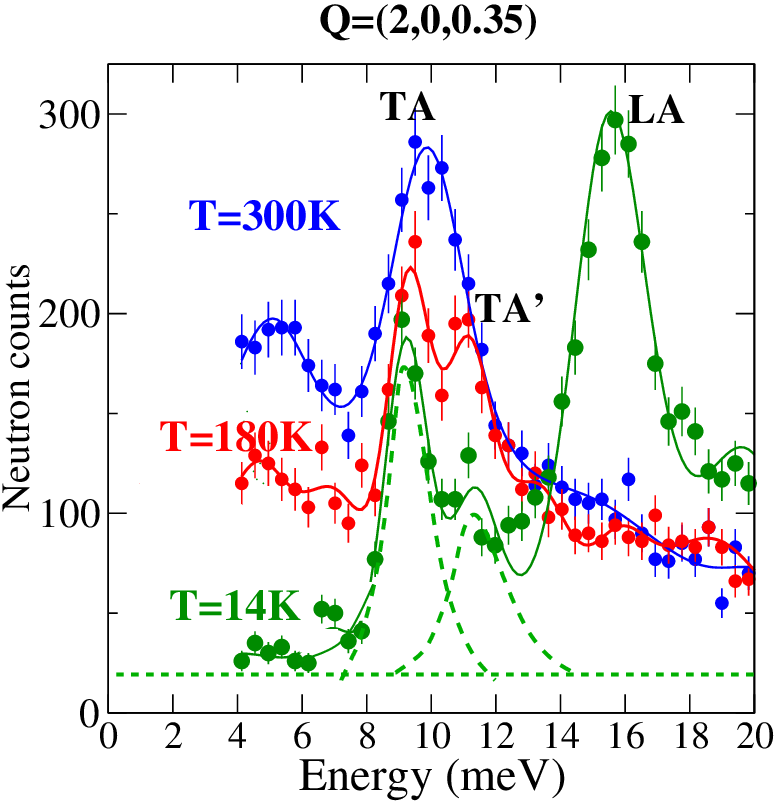}

Fig.SM-9: Examples of raw data obtained in transverse configuration at $Q=(2,0,0.35)$ for three temperatures T=14K ($T<T_{co}$) 180K ($T>T_{co}$) and 300K (($T>T_{co}$). At 300K the TA branch exhibits a large energy linewidth which splits in the two modes TA and TA' by decreasing temperature (T=180K). The most salient feature is the change in intensity of the LA branch observed in the transverse configuration at $T<T_{co}$. This observation which occurs beyond q=0.25 corresponds to the coupling between the magnetic and the acoustic phonon excitation on the scale of the orbital polarons.
\end{figure}

The raw data of Fig.SM-9 illustrate the evolution of the TA' and LA branches obtained in the transverse configuration as a function of temperature. At 300K the large linewidth of the TA branch prevents one to observe the TA' branch which  becomes measurable at T=180K only. The most significant feature is the large increase in the intensity of the LA branch measured in the transverse configuration for q values larger than $q=0.25$. This increase occurs at $T<T_{co}$ as the ferromagnetic orbital polarons order, which induces the coincidence of the ferromagnetic and elastic excitations (see Fig.SM-5).

\begin{figure}[H]
\includegraphics[width=8cm]{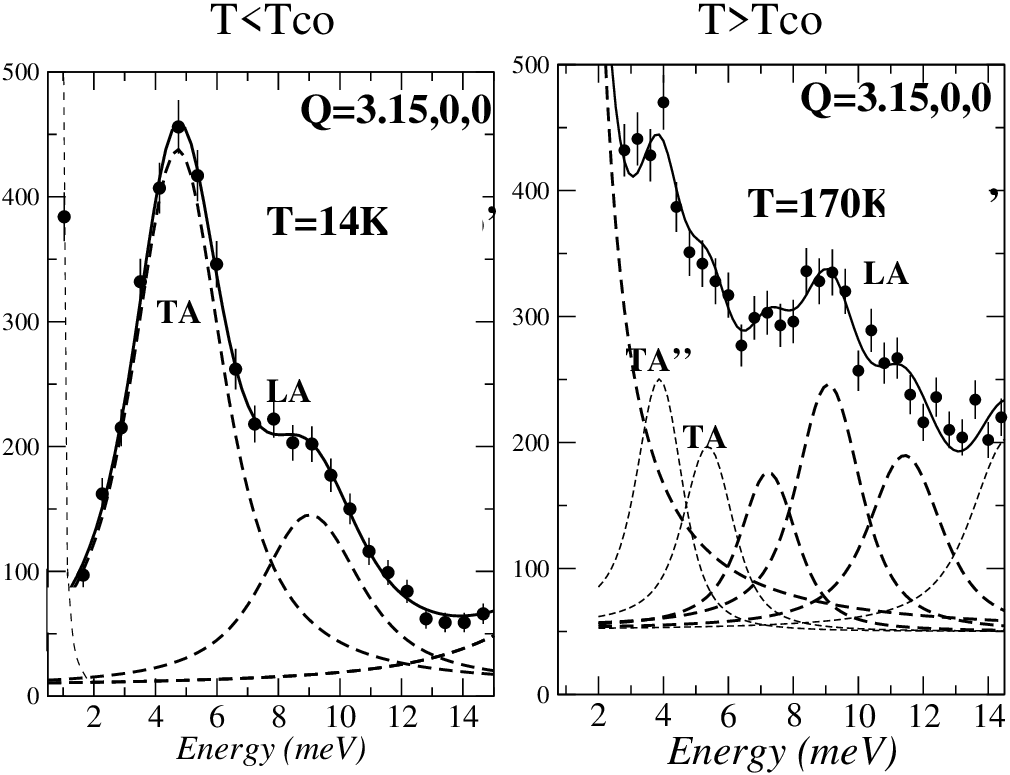}
Fig.SM-10: Raw data obtained in longitudinal configuration at $q=0.15$ rlu. Left panel: at 14K ($T<T_{co}$) the LA mode is unique for the three [100]+[010]+[001] directions, superposed because of twinning. Right panel: at T=170K ( $T>T_{co}$) three distinct modes are observed attributed to the three twinned domains of the pseudocubic structure.
\end{figure}
 
Fig.SM-10 displays raw spectra measured at $q=0.15$ rlu in the longitudinal configuration at $T<T_{co}$ (left panel) and $T>T_{co}$ (right panel). It illustrates the change of the LA branch from a single mode (left panel, $T<T_{co}$) to three modes (right panel $T>T_{co}$) for $q<0.3$ rlu, on a scale larger than the $q_{min}$ value, characteristic of the orbital bipolarons. The three modes are attributed to the three twined domains of the pseudo-cubic structure.  We also emphasize that the TA branch can be observed in the longitudinal configuration, whereas the intensity of the LA branch does not diverge. These observations contradict the usual characteristics of the longitudinal and transverse acoustic branches in a pseudocubic structure and, therefore, stress the peculiarities of these acoustic phonon branches. \\

\begin{figure}[H]
\includegraphics[width=8.5cm, scale=1.2]{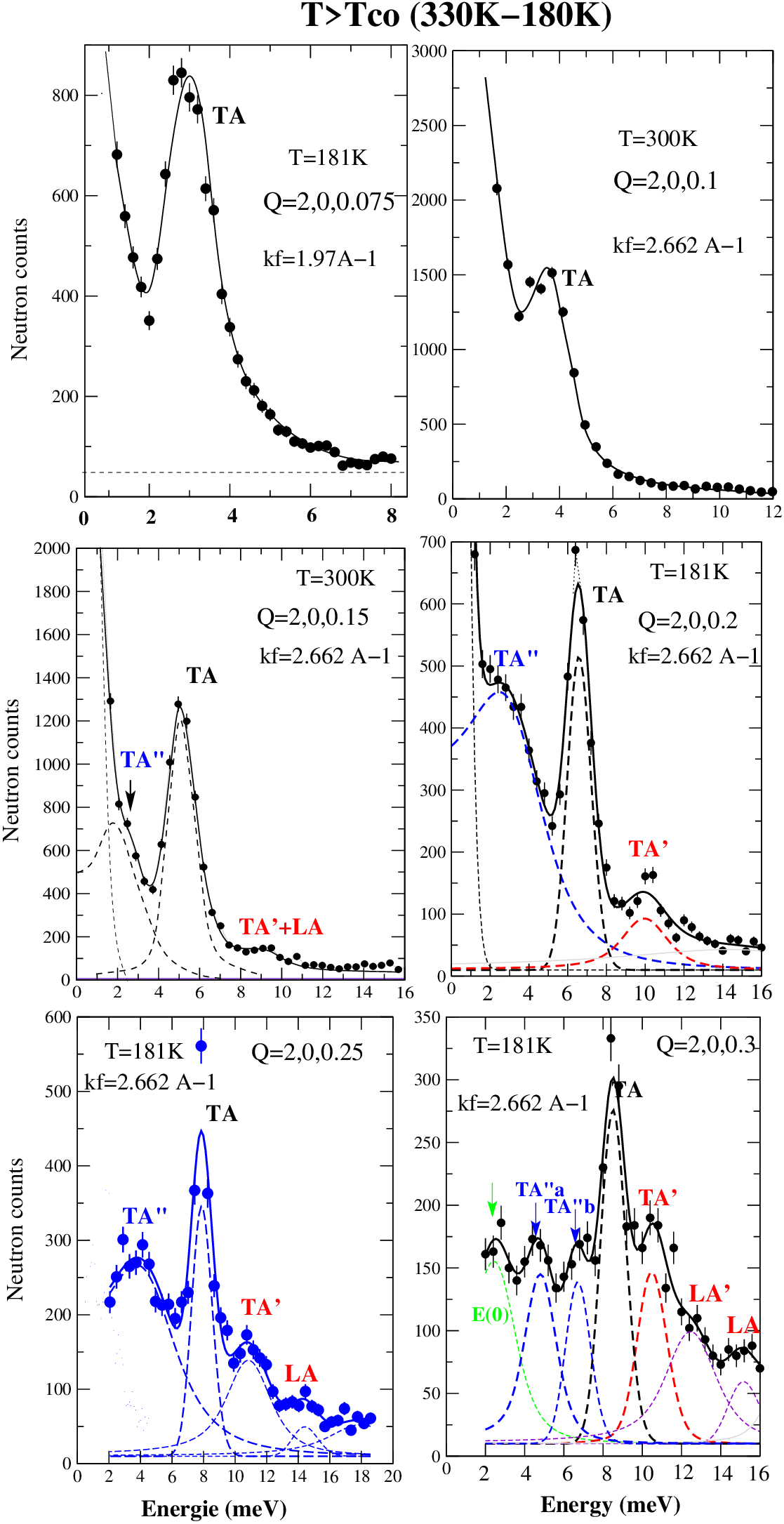}
Fig.SM-11:  Examples of raw data obtained in transverse configuration at six values of $q$ (0.075, 0.1, 0.15, 0.2, 0.25 and 0.3 in rlu units) for $T > T_{co}$ ($T=180K$ or $T=300K$), using either $k_f$=1.97 $A^{-1}$ or $2.662$ $A^{-1}$ final neutron wave vectors. Two additional excitations labeled TA" and TA' are observed in addition to the excitation of the main TA branch. For $q\ge 0.3$, the TA" excitation splits into two ones. At lowest energy, the E(0) value can be attributed to the binding energy of the polaron to the lattice. It has been also observed in the magnetic spectrum reported in Fig.6-b in the main text and fixed here in the fitting procedure. The background of $\approx 15-20$ counts has been obtained from the experiments performed at T=14K displayed in Fig.SM-4.
\end{figure}

Fig.SM-11 shows the raw data obtained in the transverse configuration at $T > T_{Tco}$ (T=181K and 300K) for six values of $q$. They illustrate the observation of the two branches TA'(q) and TA"(q) beside the main TA(q) branch.  From these raw data, one determines the value $q_{min}$ = 0.15 + / - 0.05 for the TA" branch. This determination of the $q_{min}$ value cannot be made for the TA'(q) branch due to contamination of LA(q) at $q\le 0.15 $ rlu.

\begin{figure}[H]
\includegraphics[width=8cm]{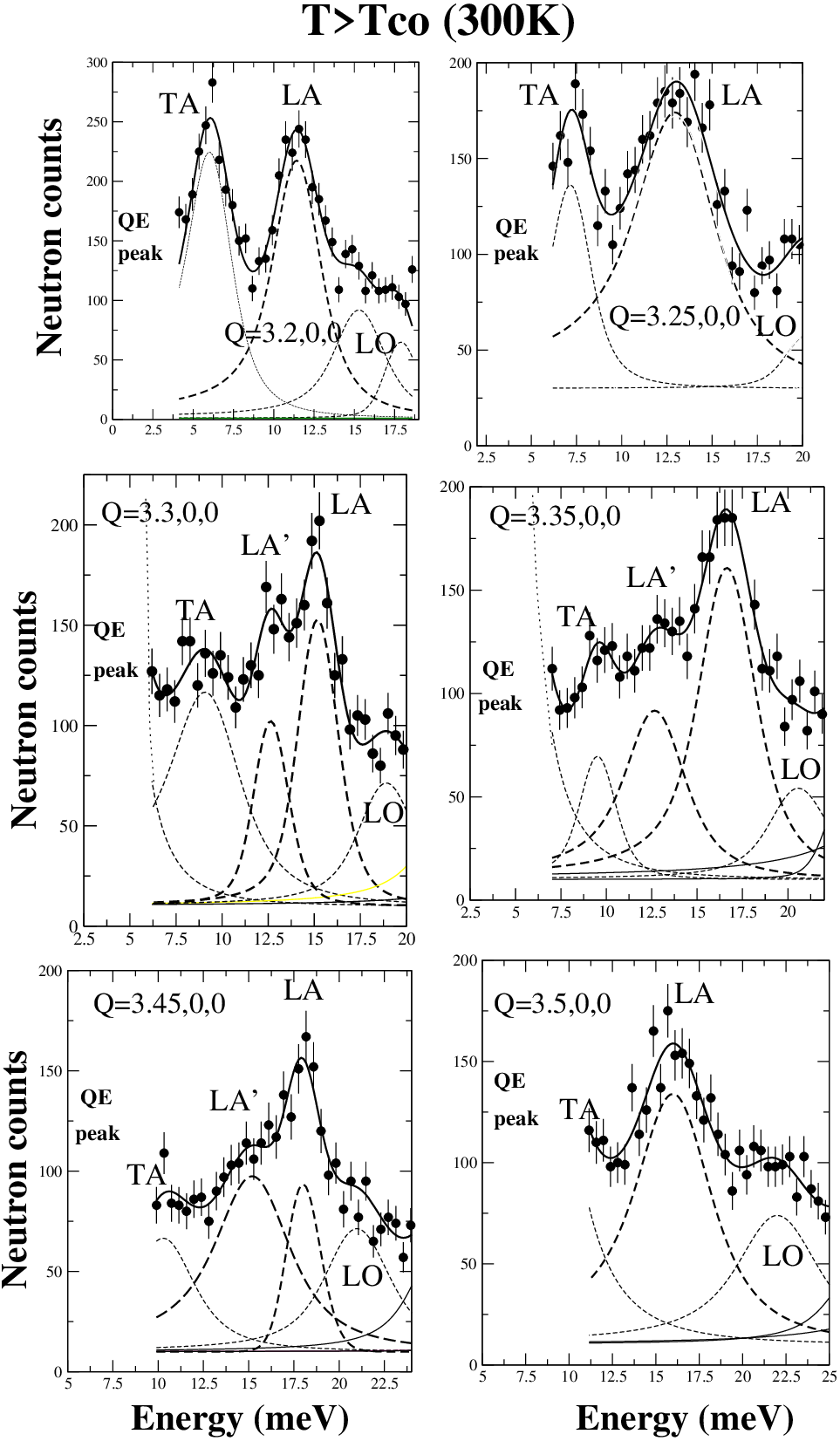}
Fig.SM-12:  Raw data obtained in longitudinal configuration at $T>T_{co}$ (T=300K) at six values of $q$. 
\end{figure}
Fig.SM-12 shows the raw data obtained in the longitudinal configuration at $T>T_{co}$ (T=300K). They illustrate the evolution of longitudinal acoustic excitations with $q$. The single peak observed at $q=0.2$ rlu broadens at $q=0.25$ rlu. Then it splits into a double peak attributed to the LA and LA' branches, as reported at $q\ge 0.3$, $q=0.35$ and $q=0.45$. At $q=0.5$ rlu, the LA' and LA excitations are no longer separated within our energy resolution.

\begin{figure}[H]
\includegraphics[width=8cm]{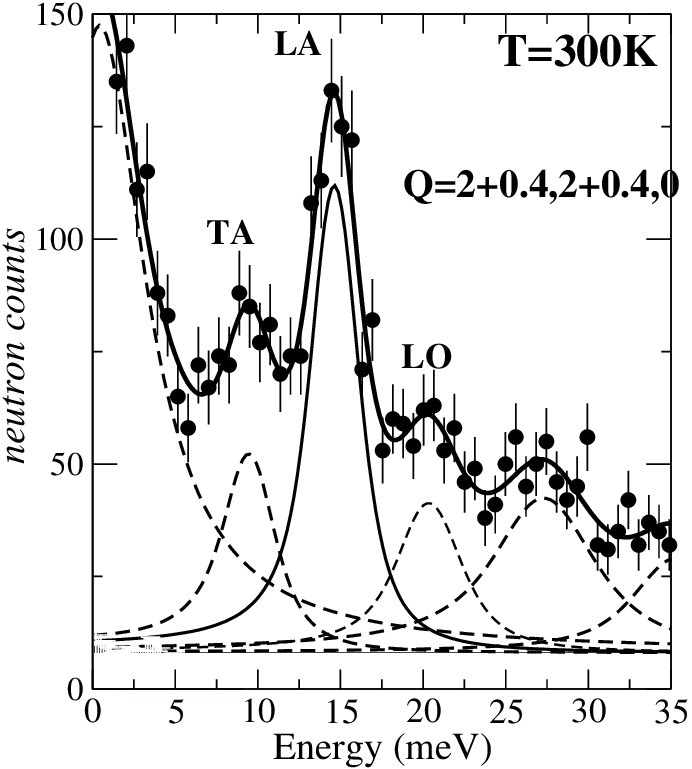}
Fig.SM-13:  Raw data of the LA excitation energies obtained at $Q=(2.4,2.4,0)$ along [110] and T=300K. Unlike along [100], no additional excitation energy can be observed close to the LA branch.
\end{figure}

Fig.SM-13 displays raw data of the phonon spectrum measured at $q=0.4$ rlu (T=300K) along [110], corresponding to the data of Fig.1-b in the main text. In contrast to the [100] direction, a single LA mode is observed, well defined. We conclude that the additional phonon excitation characteristics of bipolarons are specific of the MnO bond directions.\\\\



{\bf References (additional)}

86. G. Biotteau, M. Hennion, M. Moussa, F. Rodriguez-Carvajal, L. Pinsard, A. Revcolevschi, Y. M. Mukovsky and D. Shulyatev Phys. Rev. B {\bf 64}, 104421 (2001); and F. Moussa et M. Hennion in: Colossal Magnetoresistive Manganites, ed. T. Chatterji, Kluwer Academic Publishers Dordrecht Netherlands (2002)




87. M. Hennion and F. Moussa  New Journal of Physics {\bf 7}, 84 (2005)







88. A. Urushibara, Y. Moritomo, T. Arima, A. Asamitsu, G. Kido and Y. Tokura Phys. Rev. B {\bf 51}, 14101 (1995)











\begin{thebibliography}{99}
\bibitem{Note} This mismatch is much too large to be related to any lattice distortion unlike it was written in Ref.\onlinecite{Hennion19}
\end{thebibliography}
\end{document}